\documentclass[a4paper,11pt]{article}
 
\usepackage{amsmath}
\allowdisplaybreaks
\usepackage{amssymb} 
\usepackage{array} 
\usepackage{authblk}
\usepackage[british]{babel}
\usepackage{fullpage}
\usepackage{graphicx}
\usepackage[colorlinks=true,linkcolor=red,urlcolor=blue, citecolor=blue,anchorcolor=blue]{hyperref}
\usepackage[version=4]{mhchem}
\usepackage[numbers,sort&compress]{natbib}
\usepackage{setspace}
\usepackage{tikz}
\usepackage{xcolor} 

\begin{document}
 
\onehalfspacing
 
\title{\large \bf Clustering versus sorting: a mass-conserving reaction-diffusion model \\of planar polarity puncta}
\author[1,2]{Sk Raj Hossein}
\author[1,3]{Eman Alwani}
\author[4]{Andreas Buttensch\"on}
\author[1,*]{Alexander G. Fletcher}
\date{}
\affil[1]{School of Mathematical and Physical Sciences, University of Sheffield, Hicks Building, Hounsfield Road, Sheffield, S3 7RH, UK}
\affil[2]{School of Biological Sciences, University of Edinburgh, Mayfield Road, Darwin Building King's Buildings, Edinburgh, EH9 3JR, UK}
\affil[3]{Mathematics Department, Faculty of Sciences, Umm Al-Qura University, P.O. Box 14035, Makkah, 21955, Saudi Arabia}
\affil[4]{Department of Mathematics and Statistics, University of Massachusetts Amherst, 710 N Pleasant St, Amherst, MA 01003, USA}
\affil[*]{Corresponding author: a.g.fletcher@sheffield.ac.uk}
 
\maketitle

\section*{Abstract}
 
Planar cell polarity is preceded by the clustering of polarity proteins into discrete, low-turnover membrane subdomains (puncta), yet the minimal interactions that nucleate puncta, set their number, and segregate opposite orientations remain unclear. 
We address these questions with a mass-conserving reaction-diffusion model in which two diffusible monomers bind reversibly across a cell-cell junction into trans-complexes of two orientations, with feedback entering only through concentration-dependent rates. 
Above a critical density, the uniform state undergoes a long-wavelength mass-redistribution instability rather than a finite-wavelength Turing bifurcation. 
The form of the feedback then selects between two morphologies: broad mesas fixed by a Maxwell construction under saturating feedback, and narrow mass-limited spikes under unbounded feedback. 
For spikes we obtain closed-form expressions exhibiting a clean separation of amplitude (mass and feedback), width (the complex diffusion length), and spacing (the monomer screening length). 
Within a fast-monomer reduction we prove, for any number and arrangement of puncta, that like-oriented arrays coarsen, so multiplicity is metastable and kinetically determined. 
The second monomer reservoir introduces a second screening length that rate-limits competition by the harmonic mean of the monomer diffusivities, and the spectrum of a punctum remains free of oscillatory (``blinking") instabilities throughout. 
Finally, sign-definite cross-modulation of turnover converts clustering into orientation sorting -- the mutual exclusion of orientations along a single contact: mass redistribution sets puncta number, and the sign of the cross-coupling determines whether orientations segregate. 
Puncta number and orientation sorting are thus governed by mathematically separable ingredients.

\clearpage
\section{Introduction} \label{sec:introduction}
 
The patterning of multicellular tissues relies on self-organisation that emerges from iterative interactions across scales, from molecules to cells to tissues~\citep{karsenti2008self, werner2017self}. 
At the molecular scale, interactions through reaction-diffusion networks can break symmetry and polarise cells~\citep{gierer1972theory, li2010symmetry}. 
A recurrent motif in this process is the concentration of proteins into discrete membrane subdomains, or \emph{puncta}. These are localised clusters where the participating proteins turn over more slowly than elsewhere.
In this paper we are concerned exclusively with such molecular-scale patterning -- the formation, number and arrangement of puncta along a single cell--cell contact -- rather than with cell- or tissue-scale patterning. 
The domain throughout is one contact, so what we call \emph{sorting} is the mutual exclusion of oppositely oriented complexes \emph{along} that contact. 
Cell-scale planar polarity is a distinct statement -- that one orientation prevails at one edge of a cell and the opposite orientation at the opposite edge -- and requires coupling between the several contacts of a cell, together with a bias that selects which prevails where. 
Neither ingredient is present here.
 
Planar cell polarity provides a particularly clear instance of this motif.
In the \emph{Drosophila} wing, both the core and the Fat--Dachsous (Ft--Ds) pathways concentrate their components into puncta of locally aligned, asymmetric intercellular complexes at proximo-distal cell junctions~\citep{strutt2011dynamics, strutt2016robust, brittle2012planar}. 
Within these puncta, the complexes are markedly more stable and turn over more slowly than in the intervening non-punctate membrane~\citep{strutt2011dynamics, strutt2016robust}. 
These puncta appear functionally important for generating cellular asymmetry~\citep{strutt2011dynamics}. 
Yet the mechanisms that nucleate stable puncta, that set how many form along a cell--cell junction, and that sort complexes of opposite orientation to opposite cell edges remain incompletely understood~\citep{strutt2016robust, aigouy2010cell}.
 
At the molecular level, the core pathway comprises a small set of conserved proteins that partition into intercellular complexes of opposite orientation on the two sides of each junction~\citep{strutt2011dynamics, strutt2016robust}. 
These include the transmembrane proteins Frizzled, Flamingo, and Van Gogh/Strabismus, alongside the cytoplasmic proteins Dishevelled, Diego, and Prickle.
Cooperative, stoichiometrically flexible (``signalosome-like'') assembly stabilises like-oriented complexes and slows their turnover~\citep{strutt2008differential, strutt2016robust}. Competitive, phosphorylation-regulated interactions -- notably the destabilisation of oppositely oriented complexes by Prickle, modulated by Casein Kinase~I$\varepsilon$ -- instead sort the two orientations to opposite cell edges~\citep{warrington2017dual, cho2015clustering, strutt2019reciprocal}. 
These two ingredients, cooperative clustering and competitive cross-destabilisation, motivate the feedback terms in the model developed below.
 
These observations motivate a set of generic questions about puncta that are, in principle, independent of the precise molecular identities involved. 
What minimal interactions are sufficient to establish a \emph{stable} punctum, as opposed to a transient fluctuation? 
When does a cell-cell junction support a single stable punctum versus several, and what selects that number? 
And how do the size, shape, spacing and stability of puncta depend on the underlying reaction and diffusion parameters? 
A closely related question is whether the same machinery that drives clustering also drives \emph{sorting} -- the mutual exclusion of oppositely oriented complexes along a contact, which is thought to underlie the amplification of asymmetry at the molecular scale. 
Experimentally, planar polarity is thought to combine a weak preference for asymmetric complex formation~\citep{strutt2008differential} with feedback that mutually destabilises oppositely oriented complexes. 
This feedback is regulated by phosphorylation of complex components~\citep{warrington2017dual, strutt2019reciprocal}, and earlier modelling suggests that a weak formation bias together with sorting feedback can suffice~\citep{fisher2019modelling}.
 
To address these questions independently of specific pathway details, we adopt an idealised description. 
Here, the molecular cast is reduced to two generic species binding across a cell-cell junction to form complexes of two orientations.
While this minimal setting sacrifices molecular realism for analytical tractability, it isolates the key ingredients common to both the core and Ft-Ds pathways: mass-conserving binding, clustering feedback, and a cross-modulation of turnover. 
This economy allows us to investigate which questions can be answered generically.
 
\subsection{Previous modelling, aims and contributions} \label{sec:modelling_background}

Existing models of planar polarity have largely been built to explain alignment and coordination at the cell and tissue scales. 
Reaction kinetic and computational models of the core pathway reproduce the asymmetric build-up of complexes, domineering non-autonomy of clones, and the propagation of polarity between cells~\citep{amonlirdviman2005mathematical, burak2009order, legarrec2006establishment, hazelwood2013functional, fischer2013persistent}. 
Separately, our own earlier modelling indicates that a weak bias towards asymmetric complex formation, combined with sorting feedback, can align complexes within a cell~\citep{fisher2019modelling}.
These models typically treat each junction as well mixed, or otherwise neglect spatial variation at the subcellular level. 
By construction they do not resolve the puncta themselves, and so cannot address what sets a punctum's size, how many puncta a cell--cell junction supports, or how these depend on diffusion and feedback.
 
The mathematical ingredients needed to resolve such localised structures are, however, well developed in the pattern-formation literature. 
Clustering of a diffusible species under cooperative feedback is the classical setting of Turing and activator--substrate (Gierer--Meinhardt) systems~\citep{gierer1972theory}, whose localised spike solutions and their stability have been analysed in detail~\citep{iron2001stability}, including in the near-shadow, mesa-forming limit~\citep{mckay2012stability}. 
When the dynamics instead conserve total mass, the relevant mechanism is wave pinning~\citep{mori2008wave}. 

Much of the qualitative behaviour of localised states in that conserved setting is by now well understood, and we take it as our starting point rather than as a result. 
\citet{otsuji2007mass} showed that mass-conserved reaction--diffusion systems generically select a \emph{single} peak, the multi-peak states being transients~\citep{ishihara2007transient}, and \citet{brauns2021wavelength} established that coarsening in two-component systems of this kind is generically uninterrupted. 
The onset, coarsening and interface dynamics have since been organised into a phase-space and mass-redistribution framework~\citep{brauns2020phase, frey2026pattern}, complemented by bifurcation analyses of conserved polarity models~\citep{buttenschon2022cell}. 
In the cell-polarity setting, \citet{chiou2018principles, chiou2021how} showed that competition between polarity sites slows sharply as the conserved load increases, so that multiple sites coexist on biologically relevant timescales -- metastability rather than stability. 
We recover all of these behaviours below, and indicate where they are inherited. 

Two features of the present problem lie outside that body of work. 
The first is the pair of reservoirs: these analyses concern a \emph{single} conserved species, or a single active/inactive pair, whereas a \emph{trans}-complex draws on two independently conserved reservoirs, one presented on each face of the junction. 
The second is sorting: they treat clustering in isolation from the segregation of oppositely oriented structures.
 
The present work brings these two strands together in the specific setting of polarity puncta. 
Its organising question is what changes when the conserved object is a \emph{trans}-complex: two independently conserved monomer reservoirs, one presented on each face of the junction, feeding a single population of complexes. 
Our contribution is threefold, and we distinguish at the outset between what the heterodimer inherits from the single-reservoir theory and what is genuinely new to it.

First, we show that the mass-conserving heterodimer patterns by the same long-wavelength, mass-redistribution route as its single-reservoir counterparts, and we assemble a closed-form description of a single punctum. 
This is a matter of setting rather than of novelty: the inner problem reduces to the classical spike equation~\citep{gierer1972theory, iron2001stability}, with the \emph{product} of the two reservoirs in place of the square of one, so the mathematics is standard. 
The reduction does, however, separate the three observables of a punctum. 
Amplitude, a universal width and spacing each trace to a different model ingredient, and the heterodimer results below build on that separation.

Second, and centrally, the two reservoirs endow the puncta with a \emph{second} screening length that has no single-reservoir analogue. 
We derive in closed form, with no adjustable constant, that competition between puncta is rate-limited by the \emph{harmonic mean} of the two monomer diffusivities, and we confirm the law numerically across the $(D_{a},D_{b})$ plane. 
This is the one place where the heterodimer departs qualitatively from a single-monomer model. 
Alongside it we establish that every like-oriented array is unstable and coarsens. 
This is known for single-reservoir systems~\citep{otsuji2007mass, brauns2021wavelength}; we prove it here for any number and arrangement of puncta, by a sign-definiteness argument that uses only the decay-free structure of the monomer operator. 
The surviving number is therefore set kinetically rather than by a fixed wavelength~\citep{chiou2021how}, and the conserved kinetics forbid oscillatory (``blinking'') puncta.

Third, we identify the minimal addition that turns clustering into sorting, namely a sign-definite cross-modulation of complex turnover, and show analytically that it selects an antisymmetric mode by which co-located puncta of opposite orientation segregate along the contact. 
That mutual destabilisation should produce mutual exclusion is unsurprising in outline; what the analysis supplies is the quantitative content. 
The criterion is exact and sign-definite at linear order for an arbitrary clustering feedback, the segregation rate of fully formed puncta is $\tfrac{8}{7}\mathcal{V}'c_{\max}$, and the drift coefficient governing a separated pair varies by more than a factor of four between contact and the far field, so that no constant-gradient approximation reproduces both limits. 
This is a claim about the molecular scale alone: it delivers mutual exclusion of the two orientations, not the cell-scale asymmetry that additionally requires the contacts of a cell to be coupled. 
Clustering and sorting are thereby controlled by separate, identifiable ingredients.
 
Throughout, the aim is to determine analytically under what conditions stable puncta can arise -- as a Turing instability, a wave-pinning instability, or both. The remainder of the paper is organised as follows. 
Section~\ref{PCP_model} introduces the reaction--diffusion model and its non-dimensionalisation. Section~\ref{sec:onset} determines when the spatially uniform state loses stability, identifies the instability as a long-wavelength, mass-redistribution mode, and interprets it through the well-mixed bistable kinetics. 
Section~\ref{sec:single} analyses the structure of an individual punctum, distinguishing the broad mesas produced by saturating feedback from the narrow spikes produced by unbounded feedback. 
Section~\ref{sec:arrays} treats arrays of puncta -- their competition and coarsening, the effect of asymmetric monomer diffusion, and the absence of oscillatory (``blinking'') states. 
Section~\ref{PCP_polarity} shows how a sign-definite cross-modulation of complex turnover turns clustering into sorting. 
We discuss the biological implications and limitations of the model in Section~\ref{sec:discussion}.
 
\section{The model} \label{PCP_model}
 
\subsection{Governing equations}
 
\begin{figure}[htbp]
\centering
\begin{tikzpicture}[>=stealth, font=\small, xscale=1.25]
  \fill[black!6] (0,0) rectangle (10,2.2);
  \draw[very thick] (0,2.2) -- (10,2.2);
  \draw[very thick] (0,0) -- (10,0);
  \node[anchor=south west] at (0,2.27) {membrane of cell\ $i$};
  \node[anchor=north west] at (0,-0.07) {membrane of cell\ $i{+}1$};
  \draw[dashed] (0,-0.75) -- (0,2.85);
  \draw[dashed] (10,-0.75) -- (10,2.85);
  \node[below] at (0,-0.75) {$X=0$};
  \node[below] at (10,-0.75) {$X=L$};
  \draw[<->] (0,-0.65) -- node[below=1pt] {junction, coordinate $X$} (10,-0.65);
  \node[align=center] at (-0.05,1.10) {\scriptsize zero\\[-2pt]\scriptsize flux};
  \node[align=center] at (10.05,1.10) {\scriptsize zero\\[-2pt]\scriptsize flux};
  \foreach \x in {1.2,8.0}{\fill[blue!70] (\x,1.78) circle (0.10);}
  \foreach \x in {2.4,9.0}{\fill[orange!85!black] (\x,1.78) circle (0.10);}
  \foreach \x in {1.6,7.4}{\fill[blue!70] (\x,0.42) circle (0.10);}
  \foreach \x in {3.0,8.6}{\fill[orange!85!black] (\x,0.42) circle (0.10);}
  \node[blue!70!black]        at (1.2,2.02) {\scriptsize $A$};
  \node[orange!85!black]      at (2.4,2.02) {\scriptsize $B$};
  \node[blue!70!black]        at (7.4,0.16) {\scriptsize $A^{\dagger}$};
  \node[orange!85!black]      at (8.6,0.16) {\scriptsize $B^{\dagger}$};
  \draw[->, blue!70, thick] (2.4,1.78) -- (3.5,1.78);
  \draw[->, blue!70, thick] (8.0,1.78) -- (6.9,1.78);
  \node[blue!70!black] at (5.2,1.78) {\scriptsize fast: $D_{A},D_{B}$};
  \foreach \x in {4.7,5.0,5.3,4.85,5.15}{\fill[violet!75] (\x,1.02) circle (0.13);}
  \node[violet!75!black] at (5.0,1.42) {\scriptsize $C,\,C^{\dagger}$ (punctum)};
  \draw[->, violet!75, thick] (5.6,1.02) -- (6.0,1.02);
  \draw[->, violet!75, thick] (4.4,1.02) -- (4.0,1.02);
  \node[violet!75!black] at (5.0,0.62) {\scriptsize slow: $D_{C}\ll D_{A},D_{B}$};
  \node[align=center] at (5.0,0.18)
    {\scriptsize $C=A\!\cdot\!B^{\dagger}$,\quad $C^{\dagger}=A^{\dagger}\!\cdot\!B$};
  \draw[<->] (3.6,1.02) .. controls (4.1,1.45) .. (4.55,1.10);
  \node at (3.4,1.42) {\scriptsize $k,\,v$};
\end{tikzpicture}
\caption{\textbf{Schematic of the model.} 
A cell--cell junction is idealised as a one-dimensional domain $0<X<L$ between two apposed membranes. 
Unbound monomers $A$ and $B$ are presented on the face of cell $i$, and their opposite-face partners $A^{\dagger},B^{\dagger}$ on the face of cell $i{+}1$; colour tracks species identity, so that $A$ and $A^{\dagger}$ (blue) are the same molecule on the two sides, as are $B$ and $B^{\dagger}$ (orange). 
The monomers diffuse rapidly along the junction and bind reversibly in \emph{trans} (rates $k,v$) into slowly diffusing complexes of two orientations, each pairing one monomer from each face -- $C=A\!\cdot\!B^{\dagger}$ and $C^{\dagger}=A^{\dagger}\!\cdot\!B$, as in \eqref{eq:reaction1}--\eqref{eq:reaction2} -- which cluster into puncta. 
The faster diffusion of the monomers relative to the complex, together with mass conservation on the finite domain (zero-flux ends), are the ingredients underlying the patterning analysed below.}
\label{fig:schematic}
\end{figure}
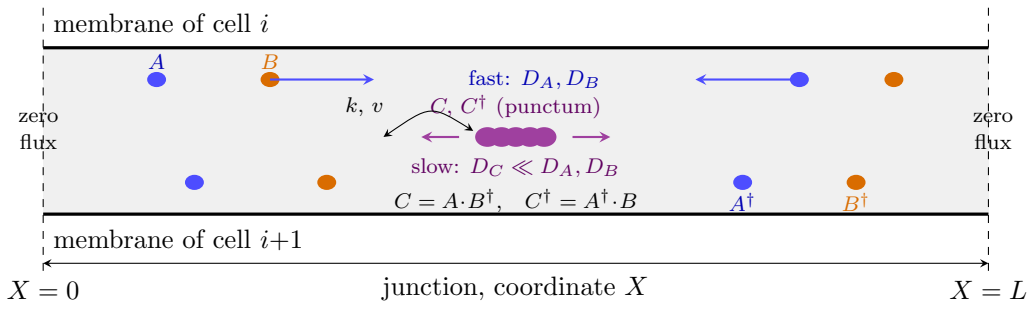

Rather than commit to the molecular identities of specific core or Ft--Ds components, we deliberately model an idealised polarity complex built from two generic species. 
This keeps the model agnostic about the precise biochemistry -- whose details differ between pathways and remain partly unresolved -- while retaining the features common to both, namely reversible \emph{trans}-binding, mass conservation and concentration-dependent feedback, and it is this economy of ingredients that makes the analysis below tractable.
Let $A(x, t)$ and $A^{\dagger}(x, t)$, $B(x, t)$ and $B^{\dagger}(x, t)$ denote the concentrations of unbound molecules $A$ and $B$ on a cell junction at position $0 < x < L$ and time $t > 0$. 
Throughout, a dagger ($\dagger$) labels the species and complexes presented on the opposite face of the junction, so that $A$ and $A^{\dagger}$ are the same molecule on the two sides. 
We treat the unbound species as freely diffusing monomers. 
This is an idealisation. 
In the core pathway and planar polarity more generally, unbound proteins may be oligomeric or membrane-tethered. 
However, representing them as single diffusing species retains the essential mass-conserving binding kinetics while keeping the analysis tractable.
Let $C(x, t)$ and $C^{\dagger}(x, t)$ denote the concentrations of the \emph{trans}-complexes formed across the junction. 
Each complex is a heterodimer of one unbound $A$ on one face of the junction and one unbound $B$ on the opposite face, formed reversibly with forward rate constant $k$ and reverse rate constant $v$,
\begin{align}
A + B^{\dagger} & \ce{<=>} C, \label{eq:reaction1} \\
A^{\dagger} + B & \ce{<=>} C^{\dagger}. \label{eq:reaction2}
\end{align}
We assume that unbound $A$ and $B$ on either side of the cell junction can diffuse along the junction with positive diffusion coefficient $D_{A}$ and $D_{B}$, respectively. 
Similarly, the complexes diffuse much more slowly than the unbound molecules $A$ and $B$, with a non-negative coefficient $D_{C}$. 
A schematic of the model geometry and its key assumptions is given in Figure~\ref{fig:schematic}.

We assume throughout that $D_{A}$, $D_{B}$ and $D_{C}$ are constant, while $k$ and $v$ may be constant or depend on the local complex concentration.
This leads to the reaction-diffusion system
\begin{align}
A_{T} & = D_{A} A_{XX} - k A B^{\dagger} + v C, \label{A:eq} \\
A^{\dagger}_{T} & = D_{A} A^{\dagger}_{XX} - k  A^{\dagger} B + v C^{\dagger}, \label{Adagger:eq}  \\
B_{T} & = D_{B} B_{XX} - k  A^{\dagger} B + v C^{\dagger}, \label{B:eq}  \\
B^{\dagger}_{T} & = D_{B} B^{\dagger}_{XX} - k A B^{\dagger} + v C, \label{Bdagger:eq} \\
C_{T} & = D_{C} C_{XX} + k A B^{\dagger} - v C, \label{C:eq} \\
C^{\dagger}_{T} & = D_{C} C^{\dagger}_{XX} + k A^{\dagger} B - v C^{\dagger},\label{Cdagger:eq} 
\end{align}
for $0 < X < L$ and $T > 0$. 
 
We model a short, representative stretch of a much longer cell--cell contact, away from the tricellular vertices where junctions meet. 
On this scale there is no preferred direction of net transport into or out of the segment. 
We therefore impose zero-flux (no-flux) boundary conditions at $X = 0$ and $X = L$,
\begin{align}
A_{X}(0,T) = A_{X}(L, T) = A^{\dagger}_{X}(0, T) = A^{\dagger}_{X}(L, T) & = 0 \nonumber \\
B_{X}(0, T) = B_{X}(L, T) = B^{\dagger}_{X}(0,T)  = B^{\dagger}_{X}(L, T) & = 0, \nonumber \\ 
C_{X}(0, T) = C_{X}(L, T) = C^{\dagger}_{X}(0, T) = C^{\dagger}_{X}(L, T) & = 0, \label{cont-no-flux_BCs}
\end{align}
and initial conditions that are assumed to be close to spatially uniform,
\begin{align}
A(X, 0) = A^{\dagger}(X, 0) & \approx A_{0}, \label{IC_A_and_Adagger} \\
B(X, 0) = B^{\dagger}(X, 0) & \approx B_{0}, \label{IC_B_and_Bdagger} \\
C(X, 0) = C^{\dagger}(X, 0) & = 0.\label{IC_C_and_Cdagger}
\end{align}
This system features four conserved quantities, the spatially averaged total densities
\begin{align}
A_{\mathrm{tot}} = \frac{1}{L} \int_{0}^{L} \left( A(X,T) + C(X,T) \right) dX, \label{A:cl} \\
A^{\dagger}_{\mathrm{tot}} = \frac{1}{L} \int_{0}^{L} \left( A^{\dagger}(X,T) + C^{\dagger}(X,T) \right) dX, \label{Adagger:cl} \\
B_{\mathrm{tot}} = \frac{1}{L} \int_{0}^{L} \left(B(X,T) + C^{\dagger}(X,T) \right) dX, \label{B:cl} \\
B^{\dagger}_{\mathrm{tot}} = \frac{1}{L} \int_{0}^{L} \left(B^{\dagger}(X,T) + C(X,T) \right) dX, \label{Bdagger:cl} 
\end{align}
where the values of $A_{\mathrm{tot}}$, $A^{\dagger}_{\mathrm{tot}}$, $B_{\mathrm{tot}}$ and $B^{\dagger}_{\mathrm{tot}}$ are determined by the initial conditions. 
 
We wish to know whether this model always evolves to a stable spatially uniform steady state. 
We will explore this through steady-state and linear stability analysis, and via numerical solution.
 
\subsection{Non-dimensionalisation} \label{PCP_Non_dimensionalisation}  
 
To remove the dimensions from the model~\eqref{A:eq}--\eqref{Cdagger:eq}, we write $k = k_{0} \mathcal{K}$ and $v = v_{0} \mathcal{V}$, where $k_{0}$ and $v_{0}$ set the scale of the binding and unbinding rates and $\mathcal{K}$, $\mathcal{V}$ are dimensionless. 
Matching $[k A B^{\dagger}] = [A]/[T]$ in \eqref{A:eq} gives $[k_{0}] = 1/(\text{concentration}\times\text{time})$ and $[v_{0}] = 1/\text{time}$. 
Concentrations are therefore measured against a reference concentration $s$, and we let $X = \mathcal{L}x$, $T = t/v_{0}$, $A = s a$, $A^{\dagger} = s a^{\dagger}$, $B = s b$, $B^{\dagger} = s b^{\dagger}$, $C = s c$, and $C^{\dagger} = s c^{\dagger}$, where $x$, $t$, $a$, $a^{\dagger}$, $b$, $b^{\dagger}$, $c$ and $c^{\dagger}$ are dimensionless. 
We then choose $\mathcal{L} = \sqrt{D_C}/\sqrt{v_0}$, so that the rescaled spatial domain is $0 < x < \ell$ with $\ell := L/\mathcal{L}$, and define dimensionless parameters
\begin{align}
A_{\mathrm{tot}}  = s\,a_{T}, \quad A^{\dagger}_{\mathrm{tot}} = s\,a^{\dagger}_{T}, \quad B_{\mathrm{tot}}  = s\,b_{T}, \quad B^{\dagger}_{\mathrm{tot}} = s\,b^{\dagger}_{T}, \quad \alpha  = \frac{k_{0}s}{v_{0}}, \quad D_{a} = \frac{D_{A}}{D_{C}}, \quad D_{b} = \frac{D_{B}}{D_{C}}.  \label{new_VariaParameters}
\end{align}
The reference concentration $s$ is a normalisation rather than a further physical parameter: it does not appear on its own in what follows, only through the dimensionless groups $\alpha$ and the scaled totals $a_{T}$, $a^{\dagger}_{T}$, $b_{T}$, $b^{\dagger}_{T}$, and the values quoted for these in Table~\ref{tab:params} correspond to a particular choice of it. 
Now substitute into~\eqref{A:eq}--\eqref{Cdagger:eq} to obtain the non-dimensionalised system 
\begin{align}
a_{t} & = D_{a} a_{xx} - \alpha \mathcal{K}(c) a b^{\dagger} +  \mathcal{V}(c^{\dagger}) c, \label{a**:eq} \\
a^{\dagger}_{t} & = D_{a} a^{\dagger}_{xx} - \alpha \mathcal{K}(c^{\dagger}) a^{\dagger} b + \mathcal{V}(c) c^{\dagger}, \label{adagger**:eq}  \\
b_{t} & = D_{b} b_{xx} - \alpha \mathcal{K}(c^{\dagger})  a^{\dagger} b + \mathcal{V}(c) c^{\dagger}, \label{b**:eq}  \\
b^{\dagger}_{t} & = D_{b} b^{\dagger}_{xx} - \alpha \mathcal{K}(c) a b^{\dagger} + \mathcal{V}(c^{\dagger}) c, \label{bdagger**:eq} \\
c_{t} & = c_{xx} + \alpha \mathcal{K}(c) a b^{\dagger} - \mathcal{V}(c^{\dagger}) c, \label{c**:eq}  \\
c^{\dagger}_{t} & = c^{\dagger}_{xx} + \alpha \mathcal{K}(c^{\dagger}) a^{\dagger} b - \mathcal{V}(c) c^{\dagger}, \label{cdagger**:eq} 
\end{align}
with dimensionless average total $A$ and $B$ concentrations
\begin{align}
a_{T} & = \frac{1}{\ell} \int_{0}^{\ell} \left(a(x,t) + c(x,t) \right) dx, \label{a:cl} \\
a^{\dagger}_{T} & = \frac{1}{\ell} \int_{0}^{\ell} \left( a^{\dagger}(x,t) + c^{\dagger}(x,t) \right) dx, \label{adagger:cl} \\
b_{T} & = \frac{1}{\ell} \int_{0}^{\ell} \left( b(x,t) +  c^{\dagger}(x,t) \right) dx, \label{b:cl} \\
b^{\dagger}_{T} & = \frac{1}{\ell} \int_{0}^{\ell} \left( b^{\dagger}(x,t) +  c(x,t) \right) dx. \label{bdagger:cl} 
\end{align}
 
\subsection{Feedback functions} \label{sec:feedback}
 
The feedbacks enter the model only through the dimensionless rate multipliers $\mathcal{K}(c)$ and $\mathcal{V}(c^{\dagger})$ appearing in \eqref{a**:eq}--\eqref{cdagger**:eq}, and we now specify the forms used throughout the analysis below.
 
The clustering feedback $\mathcal{K}(c)$ encodes the cooperative, ``signalosome-like'' assembly by which existing complex promotes the further binding of like-oriented complex~\citep{strutt2008differential, strutt2016robust}, so that the forward (binding) rate rises with the local complex density. 
We take $\mathcal{K}(0)=\mathcal{K}'(0)=0$, so that the empty membrane carries no spontaneous binding enhancement, and consider two representative choices that bracket the qualitative behaviour. 
The first is the \emph{unbounded} (Turing-type) feedback
\begin{equation}
\mathcal{K}(c)=c^{2}, \label{PCP_feedback_turing}
\end{equation}
in which the cooperative enhancement grows without bound; as we show below it produces narrow, mass-limited spikes (Section~\ref{PCP_spike}). 
The second is the \emph{saturating} (wave-pinning) feedback
\begin{equation}
\mathcal{K}(c)=\frac{c^{m}}{1+\rho\,c^{m}}, \label{PCP_feedback_sat}
\end{equation}
with Hill exponent $m$ and saturation parameter $\rho>0$, for which the enhancement levels off at high density; this produces broad, pinned mesas (Section~\ref{PCP_pinned}). 
We use $m=2$ throughout, and quote $\rho=0.1$ wherever the saturating form is invoked. 
Both choices share the cooperative onset $\mathcal{K}(0)=\mathcal{K}'(0)=0$ and differ only in whether the enhancement saturates, which is precisely what distinguishes the spike and mesa morphologies analysed in Section~\ref{sec:single}.
 
The unbinding modulation $\mathcal{V}(c^{\dagger})$ lets the turnover of each complex depend on the local concentration of the complex of \emph{opposite} orientation -- the competitive, phosphorylation-regulated cross-interaction attributed to Prickle~\citep{warrington2017dual, cho2015clustering, strutt2019reciprocal}. 
Unless stated otherwise we take the \emph{decoupled} choice $\mathcal{V}\equiv 1$, so that the two orientations turn over independently; this is the setting for the onset, single-punctum and array analyses of Sections~\ref{sec:onset}--\ref{sec:arrays}. 
The cross-coupling that generates sorting is introduced only in Section~\ref{PCP_polarity}, through the linear form $\mathcal{V}(c^{\dagger})=1+\mathcal{V}'\,c^{\dagger}$ with $\mathcal{V}'>0$, the slope $\mathcal{V}'$ setting the strength of the cross-modulation.
 
\subsection{Numerical methods} \label{sec:numerical_methods}
 
\paragraph{Spatial and temporal discretisation.} 
The system \eqref{a**:eq}--\eqref{cdagger**:eq} -- or, in the decoupled case $\mathcal{V}'=0$, a single triplet $(a,b^{\dagger},c)$ -- was solved on $0<x<\ell$ with the zero-flux conditions \eqref{cont-no-flux_BCs} by the method of lines. 
Space was discretised by second-order centred finite differences on a uniform cell-centred grid $x_{i}=(i-\tfrac{1}{2})\Delta x$, where $\Delta x=\ell/N_{x}$. 
The no-flux boundaries were imposed via a conservative one-sided stencil. 
Specifically, the discrete Laplacian $\mathsf{L}$ carries $-1$ rather than $-2$ on its first and last diagonal entries, so that its row and column sums vanish. 
Grid sizes were $N_{x}=160$--$480$, chosen so that the universal spike width (full width at half maximum $4\,\mathrm{arccosh}\sqrt{2}\approx3.53$; \eqref{PCP_spike_profile}) is resolved by no fewer than about fourteen, and typically twenty to thirty-five, points. 
Time was advanced by a first-order semi-implicit (IMEX) scheme, diffusion treated implicitly (backward Euler) and the reactions
explicitly, so that each field is updated as $u^{n+1}=(\mathsf{I}-\Delta t\,D\,\mathsf{L})^{-1}\!\bigl(u^{n}+\Delta t\,R(u^{n})\bigr)$, with $D$ the field's diffusivity and $R$ its reactive term. 
As the diffusivities, grid and step are fixed within a run, the constant-coefficient operators $\mathsf{I}-\Delta t\,D\,\mathsf{L}$ were
$LU$-factorised once and reused, reducing each step to a back-substitution; time steps were $\Delta t=0.002$--$0.01$. 
Since the reactive terms of a monomer and its complex are equal and opposite and the Neumann Laplacian is conservative, the discrete totals \eqref{a:cl}--\eqref{bdagger:cl} are preserved to the precision of the direct solve; all four were monitored throughout and their drift remained at machine precision ($\lesssim10^{-12}$). 
This finite-difference scheme is used for every time integration in the paper except the sorting runs of Figure~\ref{AB_sorting}(a)--(c), which are integrated spectrally as described under \emph{Analytical curves and software} below.

\paragraph{Spectral computations.}
Sections~\ref{PCP_competition}--\ref{PCP_polarity} quote closed forms obtained in asymptotic limits, and each is tested: the sorting results against the simulations just described, and the competition rates against the eigenvalue problem they came from.
Linearising about a $K$-punctum state couples the puncta only through the monomer reservoir, so \eqref{PCP_disp_spike}, and its unequal-diffusivity form \eqref{PCP_disp_asym}, are rank-one perturbations of a local operator whose determinant reduces to a scalar condition on $\lambda$.
We discretise the resolvent of that operator by an ultraspherical spectral method \citep{olver2013fast} and solve the scalar condition for the competition rate $\lambda_{\mathrm{comp}}$ (Figures~\ref{AB_competition_saturation} and~\ref{AB_tworeservoir_plane}).
Discretisation sizes, time steps and tolerances are given with each figure.

\paragraph{Initial data.} 
Two kinds of initial condition were used. 
For the onset simulations (Figure~\ref{AB_onset_numerics}), each field was initialized at its spatially uniform steady value plus a small, smooth, band-limited perturbation of amplitude $\epsilon$ (cf.\ \eqref{IC_C_and_Cdagger}). 
This uniform value corresponds to the upper, stable root of the well-mixed kinetics (the reactive nullcline \eqref{PCP_nullcline}). 
The perturbation was constructed as a sum of the first forty cosine harmonics with independent Gaussian, $1/k$-weighted coefficients and random phases, adjusted to be zero-mean and unit-amplitude before scaling by $\epsilon$. 
The two complex orientations were seeded independently, so that at $\mathcal{V}'=0$ their relative placement reflects the seed rather than any sorting
(Figure~\ref{AB_onset_numerics}). 
For the single-structure, array and sorting runs (Figures~\ref{AB_pinning_maxwell} and~\ref{AB_spike}--\ref{AB_sorting}) the fields were instead initialised with an analytic profile and relaxed to steady state. 
The pinned front was seeded with a smoothed (hyperbolic-tangent) step. 
The spike, coarsening and sorting runs were seeded with the asymptotic spike profile \eqref{PCP_spike_profile}: one or several $\mathrm{sech}^{2}$ peaks at prescribed positions, with a small ($1$--$10\%$) amplitude asymmetry where competition was studied. 
In both cases the monomer reservoirs were fixed by the conserved masses. 
Steady states were obtained by integrating to a time long compared with the slowest relaxation (up to $t\sim10^{3}$).

\paragraph{Diagnostics.} 
Competition rates $\lambda_{\mathrm{comp}}$ (Figures~\ref{AB_competition}(b) and~\ref{AB_tworeservoir}(a)) were measured from a weakly perturbed two-spike pair (a $\pm1\%$ amplitude offset) as the first $e$-folding of the difference between the left- and right-half complex masses -- the reciprocal of the time for that difference to grow by a factor $e$ -- and compared with the closed forms \eqref{PCP_rate_explicit} and \eqref{PCP_harmonic}. 
As those forms are the shadow-limit ($D_{a}\to\infty$) rates, the diffusivities for these panels were kept in the regime $\ell^{2}/D_{a}\lesssim4$ (Section~\ref{PCP_competition}). 
Puncta were counted (Figure~\ref{AB_coarsening}) by peak detection on $c(x,t)$. 
The single-spike spectra (Figure~\ref{AB_tworeservoir}(b)) were computed by relaxing to the steady single spike, assembling the dense $3N_{x}\times3N_{x}$ Jacobian of the discretised triplet about that profile, and finding all of its eigenvalues with a dense ($QR$) eigensolver; the real amplitude mode and the complex-conjugate oscillatory pairs were then identified.

\paragraph{Analytical curves and software.} 
The dispersion relation (Figure~\ref{AB_dispersion}) was evaluated as the largest real part of the three roots of the cubic \eqref{PCP_cubic}, obtained
mode by mode from its companion matrix; the spatially uniform steady states, the reactive nullclines \eqref{PCP_nullcline} (Figure~\ref{AB_nullcline}) and the Maxwell plateau and front position \eqref{PCP_maxwell}, \eqref{PCP_position} (Figure~\ref{AB_pinning_maxwell}) were found by bracketed root-finding (Brent's method), and involve no time integration. 
All computations were performed in Python with \textsc{NumPy} and \textsc{SciPy} (sparse $LU$ factorisation, Brent root-finding, dense eigensolver and peak detection) and \textsc{Matplotlib}. 
The first-order semi-implicit (IMEX) finite-difference scheme described above is used for every direct simulation except the sorting runs of Figure~\ref{AB_sorting}(a)--(c): specifically, for the onset simulations of Figure~\ref{AB_onset_numerics}, the single-structure and array simulations of Figures~\ref{AB_pinning_maxwell}--\ref{AB_competition} and~\ref{AB_coarsening}--\ref{AB_tworeservoir} (including the measured rates of Figure~\ref{AB_competition_saturation}(a), obtained with the diagnostic of Figure~\ref{AB_competition}(b)), and the drift-law measurements of Figure~\ref{AB_sorting}(d). 
Two computations instead use the open-source \textsc{funpy} spectral library~\citep{funpy}. 
First, the competition eigenvalue problems of Figures~\ref{AB_competition_saturation} and~\ref{AB_tworeservoir_plane}, for which \textsc{funpy}'s ultraspherical discretisation~\citep{olver2013fast} supplies the resolvent of \eqref{PCP_disp_spike} and \eqref{PCP_disp_asym}; these are spectral rootings of a scalar condition on $\lambda$ and involve no time integration. 
Second, the sorting runs of Figure~\ref{AB_sorting}(a)--(c), which are integrated with \textsc{funpy}'s fourth-order exponential (ETDRK4) Fourier scheme on the even extension of the domain onto $0<x<2\ell$, so that the restriction to $0<x<\ell$ satisfies the zero-flux conditions \eqref{cont-no-flux_BCs} exactly. 
A single script, which depends on \textsc{NumPy}, \textsc{SciPy} and \textsc{Matplotlib} throughout and additionally on \textsc{funpy} for Figures~\ref{AB_competition_saturation}, \ref{AB_tworeservoir_plane} and~\ref{AB_sorting}(a)--(c), regenerates every figure, with the per-figure grid size $N_{x}$, time step $\Delta t$ and integration time set therein and the physical parameters listed in Table~\ref{tab:params}. 
Code to reproduce all figures is available at \url{https://github.com/AlexFletcher/puncta}.

\begin{table}[htbp]
\centering
\footnotesize
\setlength{\tabcolsep}{4pt}
\caption{Parameters for all computed figures. 
All simulations use zero-flux boundary conditions and the dimensionless system \eqref{a**:eq}--\eqref{cdagger**:eq}; $\mathcal{V}=1$ (decoupled) except where a coupling $\mathcal{V}(c^{\dagger})$ is given.
Totals are symmetric, $n=a_{T}=a^{\dagger}_{T}=b_{T}=b^{\dagger}_{T}$, unless noted.}
\label{tab:params}
\begin{tabular}{l >{\raggedright\arraybackslash}p{1.9cm} >{\raggedright\arraybackslash}p{1.5cm} >{\raggedright\arraybackslash}p{2.5cm} c >{\raggedright\arraybackslash}p{2.0cm} >{\raggedright\arraybackslash}p{3.4cm}}
\hline
Fig. & $\mathcal{K}(c)$ & $\alpha$ & $(D_{a},D_{b})$ & $\ell$ & $n$ & System / quantity shown \\
\hline
\ref{AB_dispersion}      & $c^{2}$                 & $0.14$ & $(100,100)$, $(100,10)$ & $100$ & $3.9$ & Decoupled cubic \eqref{PCP_cubic}; $\max\operatorname{Re}\sigma(q)$ (analytical) \\
\ref{AB_nullcline}       & $c^{2}$ / sat.\ ($m{=}2$, $\rho{=}0.1$) & $0.14$ / $0.6$ &  --  &  --  & op.\ $3.9$ & Reactive nullcline \eqref{PCP_nullcline}, $c^{*}(n)$ (analytical) \\
\ref{AB_onset_numerics}  & (a,b) $c^{2}$; (c,d) sat.\ ($m{=}2$, $\rho{=}0.1$) & $0.14$ / $0.6$ & $(100,100)$, $(100,10)$ & $100$ & $3.9$ & Full six-species; $\epsilon=2.5$, shown at $t=100$ \\
\ref{AB_pinning_maxwell} & sat.\ ($m{=}2$, $\rho{=}0.1$) & $0.6$ & $(10^{5},10^{5})$ & $40$ & $3.9$ & Decoupled triplet; steady pinned front \\
\ref{AB_spike}           & $c^{2}$                 & $0.3$  & $(2000,2000)$ & $40$ & $2.5$ & Decoupled triplet; steady spike profile \\
\ref{AB_competition}     & $c^{2}$                 & $0.3$  & (a) $(50,50)$; (b) $D_{a}{=}50$, $100$, $200$, $400$ & $40$ & $4.5$ & Decoupled triplet; (a) $10\%$ seed asymmetry, (b) $\lambda_{\mathrm{comp}}(D_{a})$ \\
\ref{AB_competition_saturation} & $c^{2}$          & $0.3$  & (a) sweep $D_{a}$, $D_{b}{=}D_{a}$; (b) $(400,400)$ & $40$ & $4.5$ & Competition NLEP \eqref{PCP_disp_spike} rooted spectrally; (a) two spikes at $\ell/4,3\ell/4$, (b) $K{=}2$--$8$ equally spaced \\
\ref{AB_coarsening}      & $c^{2}$                 & $0.3$  & (a) $(80,80)$; (b) $D_{a}{=}40,80,160$ & $60$ & $3.9$ & Decoupled triplet; six-spike seed; $N(t)$ \\
\ref{AB_tworeservoir}    & $c^{2}$                 & $0.3$ & (a) eight $(D_{a},D_{b})\ge400$; (b) $D_{a}{=}100$, sweep $D_{a}/D_{b}$ & $40$ & (a) $4.5$; (b) $2.5$ & Decoupled triplet; (a) $\lambda_{\mathrm{comp}}$ vs harmonic mean, (b) single-spike spectrum \\
\ref{AB_tworeservoir_plane} & $c^{2}$              & $0.3$  & (a) $(D_{a},D_{b})$ plane; (b) sweep $D_{a}$, $D_{b}{=}100$, $400$, $1600$ & $40$ & $4.5$ ($\bar{a}{=}\bar{b}{=}0.5$) & Asymmetric competition NLEP \eqref{PCP_disp_asym} rooted spectrally; two spikes at $\ell/4,3\ell/4$; harmonic-mean law \eqref{PCP_harmonic} \\
\ref{AB_sorting}         & $c^{2}$                 & $0.3$  & $(10^{3},10^{3})$ & $40$ & $2.5$ ($c_{\max}{\approx}19.4$) & Two coupled triplets, $\mathcal{V}(c^{\dagger})=1+\mathcal{V}'c^{\dagger}$; (a--c) $\mathcal{V}'$ swept over $[-0.03,0.03]$: segregation outcome, separation $s(t)$, near-contact rate; (d) drift law \\
\hline
\end{tabular}
\end{table}

\section{Onset of patterning} \label{sec:onset}
 
\subsection{Spatially uniform steady states} \label{PCP Steady-state analysis}
 
We seek positive spatially uniform steady state solutions (SUSS) of the system
\eqref{a**:eq}--\eqref{c**:eq} of the form $a(x,t) \equiv a^{*}$,
$a^{\dagger}(x,t) \equiv a^{\dagger *}$, $b(x,t) \equiv b^{*}$,
$b^{\dagger}(x,t) \equiv b^{\dagger *}$, $c(x,t) \equiv c^{*}$, and
$c^{\dagger}(x,t) \equiv c^{\dagger *}$.
By the conservation laws \eqref{a:cl}--\eqref{bdagger:cl}, the unbound concentrations are slaved to the complex concentrations,
\begin{equation}
a^{*} = a_{T} - c^{*}, \quad
b^{\dagger *} = b^{\dagger}_{T} - c^{*}, \quad
a^{\dagger *} = a^{\dagger}_{T} - c^{\dagger *}, \quad
b^{*} = b_{T} - c^{\dagger *},
\label{PCP_slaving}
\end{equation}
so positivity of the unbound species requires $0 \le c^{*} \le \min(a_{T}, b^{\dagger}_{T})$ and $0 \le c^{\dagger *} \le \min(a^{\dagger}_{T}, b_{T})$.
Substituting \eqref{PCP_slaving} into the steady states of \eqref{c**:eq} and \eqref{cdagger**:eq}, any SUSS must satisfy
\begin{align}
-\alpha \mathcal{K}(c^{*}) (a_{T} - c^{*})( b^{\dagger}_{T} - c^{*}) + \mathcal{V}(c^{\dagger*}) c^{*} & = 0, \label{PCP_SUSS1} \\
-\alpha \mathcal{K}(c^{\dagger*}) (a^{\dagger}_{T} - c^{\dagger*})( b_{T} - c^{\dagger*}) + \mathcal{V}(c^{*}) c^{\dagger*} & = 0. \label{PCP_SUSS2}
\end{align}
In the case of no feedback ($\mathcal{K} = \mathcal{V} = 1$), equations~\eqref{PCP_SUSS1} and~\eqref{PCP_SUSS2} each reduce to a quadratic,
\begin{align}
(c^{*})^{2} - \left(a_{T} + b^{\dagger}_{T} + \frac{1}{\alpha}\right) c^{*} + a_{T} b^{\dagger}_{T} &= 0, \label{PCP_quad_c} \\
(c^{\dagger*})^{2} - \left(a^{\dagger}_{T} + b_{T} + \frac{1}{\alpha}\right) c^{\dagger*} + a^{\dagger}_{T} b_{T} &= 0. \label{PCP_quad_cdagger}
\end{align}
Each quadratic has two positive roots, since the sum and product of the roots are both positive. 
Only the smaller root is admissible. 
The left-hand side of \eqref{PCP_SUSS1} is negative at $c^{*}=0$ and positive at $c^{*}=\min(a_{T},b^{\dagger}_{T})$, so exactly one root lies in $\left(0,\min(a_{T},b^{\dagger}_{T})\right)$. 
That root is the one carrying the negative sign of the discriminant (and analogously for $c^{\dagger*}$),
\begin{align}
c^{*} &= \frac{1}{2}\left(a_{T} + b^{\dagger}_{T} + \frac{1}{\alpha} - \sqrt{\left(a_{T} + b^{\dagger}_{T} + \frac{1}{\alpha}\right)^2 - 4a_{T} b^{\dagger}_{T}} \right), \label{PCP_c_SUSS} \\
c^{\dagger*} &= \frac{1}{2}\left(a^{\dagger}_{T} + b_{T} + \frac{1}{\alpha} - \sqrt{\left(a^{\dagger}_{T} + b_{T} + \frac{1}{\alpha}\right)^2 - 4 a^{\dagger}_{T} b_{T}} \right). \label{PCP_cdagger_SUSS}
\end{align}
The corresponding unbound concentrations then follow from \eqref{PCP_slaving} as
\begin{align}
a^{*} & = \frac{1}{2}\left(a_{T} - b^{\dagger}_{T} - \frac{1}{\alpha} + \sqrt{\left(a_{T} + b^{\dagger}_{T} + \frac{1}{\alpha}\right)^2 - 4a_{T} b^{\dagger}_{T}} \right), \label{PCP_a_SUSS} \\
b^{\dagger *} & = \frac{1}{2}\left(b^{\dagger}_{T} - a_{T} - \frac{1}{\alpha} + \sqrt{\left(a_{T} + b^{\dagger}_{T} + \frac{1}{\alpha}\right)^2 - 4a_{T} b^{\dagger}_{T}} \right), \label{PCP_bdagger_SUSS} \\
a^{\dagger *} & = \frac{1}{2}\left(a^{\dagger}_{T} - b_{T} - \frac{1}{\alpha} + \sqrt{\left(a^{\dagger}_{T} + b_{T} + \frac{1}{\alpha}\right)^2 - 4 a^{\dagger}_{T} b_{T}} \right), \label{PCP_adagger_SUSS} \\
b^{*} & = \frac{1}{2}\left( b_{T} - a^{\dagger}_{T} - \frac{1}{\alpha} + \sqrt{\left(a^{\dagger}_{T} + b_{T} + \frac{1}{\alpha}\right)^2 - 4 a^{\dagger}_{T} b_{T}} \right). \label{PCP_b_SUSS}
\end{align}
 
\subsection{Linear stability analysis and the dispersion relation} \label{PCP_Linear stability analysis}
 
To determine under what conditions a SUSS can grow, we perform a linear stability analysis, following \citet{murray2001mathematical}. 
We assume arbitrarily small perturbations $\tilde{a}$, $\tilde{a}^{\dagger}$, $\tilde{b}$, $\tilde{b}^{\dagger}$, $\tilde{c}$ and $\tilde{c}^{\dagger}$ such that
$a = a^{*} + \tilde{a}$, $a^{\dagger} = a^{\dagger *} + \tilde{a}^{\dagger}$, $b = b^{*} + \tilde{b}$, $b^{\dagger} = b^{\dagger *} + \tilde{b}^{\dagger}$, $c = c^{*} + \tilde{c}$, and $c^{\dagger} = c^{\dagger *} + \tilde{c}^{\dagger}$.
Substituting into~\eqref{a**:eq}--\eqref{c**:eq} and dropping higher-order terms, we obtain the linearised system     
\begin{align}
\tilde{a}_{t} & = D_{a} \tilde{a}_{xx} - \alpha b^{\dagger *} \mathcal{K}(c^{*}) \tilde{a} - \alpha a^{*} \mathcal{K}(c^{*})\tilde{b}^{\dagger} + \left(\mathcal{V}(c^{\dagger *}) - \alpha a^{*}b^{\dagger *} \mathcal{K}' (c^{*})\right) \tilde{c} +  c^{*}\mathcal{V}'(c^{\dagger *}) \tilde{c}^{\dagger}, \label{PCP_tilde:A*} \\
\tilde{b}^{\dagger}_{t} & = D_{b} \tilde{b}^{\dagger}_{xx} - \alpha b^{\dagger *} \mathcal{K}(c^{*}) \tilde{a} - \alpha a^{*} \mathcal{K}(c^{*})\tilde{b}^{\dagger} + \left(\mathcal{V}(c^{\dagger *}) - \alpha a^{*}b^{\dagger *}  \mathcal{K}' (c^{*})\right) \tilde{c} +  c^{*}\mathcal{V}'(c^{\dagger *}) \tilde{c}^{\dagger}, \label{PCP_tilde:Bdagger*} \\
\tilde{c}_{t} & = \tilde{c}_{xx} + \alpha b^{\dagger*} \mathcal{K}(c^{*}) \tilde{a} + \alpha a^{*} \mathcal{K}(c^{*})\tilde{b}^{\dagger} - \left( \mathcal{V}(c^{\dagger *}) - \alpha a^{*}b^{\dagger *} \mathcal{K}' (c^{*}) \right) \tilde{c} -  c^{*}\mathcal{V}'(c^{\dagger *}) \tilde{c}^{\dagger}, \label{PCP_tilde:C*} \\
\tilde{a}^{\dagger}_{t} & = D_{a} \tilde{a}^{\dagger}_{xx} + c^{\dagger*}\mathcal{V}'(c^{*}) \tilde{c} - \alpha b^{*} \mathcal{K}(c^{\dagger *}) \tilde{a}^{\dagger} - \alpha a^{\dagger *}\mathcal{K}(c^{\dagger *})\tilde{b} + \left( \mathcal{V}(c^{*}) - \alpha a^{\dagger *}b^{*} \mathcal{K}' (c^{\dagger*}) \right) \tilde{c}^{\dagger}, \label{PCP_tilde:Adagger*}  \\
\tilde{b}_{t} & = D_{b} \tilde{b}_{xx} + c^{\dagger*}\mathcal{V}'(c^{*}) \tilde{c} - \alpha b^{*} \mathcal{K}(c^{\dagger *}) \tilde{a}^{\dagger} - \alpha a^{\dagger *}\mathcal{K}(c^{\dagger *})\tilde{b} + \left( \mathcal{V}(c^{*}) - \alpha a^{\dagger *}b^{*} \mathcal{K}' (c^{\dagger*}) \right) \tilde{c}^{\dagger}, \label{PCP_tilde:B*} \\
\tilde{c}^{\dagger}_{t} & = \tilde{c}^{\dagger}_{xx} - c^{\dagger*}\mathcal{V}'(c^{*}) \tilde{c} + \alpha b^{*} \mathcal{K}(c^{\dagger *}) \tilde{a}^{\dagger} + \alpha a^{\dagger *}\mathcal{K}(c^{\dagger *})\tilde{b} - \left( \mathcal{V}(c^{*}) - \alpha a^{\dagger *}b^{*} \mathcal{K}' (c^{\dagger*}) \right) \tilde{c}^{\dagger}. \label{PCP_tilde:Cdagger*}
\end{align}
After removing tildes, the above system can be written in matrix form as
\begin{equation}
\mathbf{U}_{t} = D \mathbf{U}_{xx} + J \mathbf{U}, \label{PCP_eq:linearised_system}
\end{equation}
where
\begin{equation}
D =  
\begin{pmatrix}
D_a & 0 & 0 & 0 & 0 & 0 \\
0 & D_b & 0 & 0 & 0 & 0 \\
0 & 0 & 1 & 0 & 0 & 0 \\
0 & 0 & 0 & D_a & 0 & 0 \\
0 & 0 & 0 & 0 & D_b & 0 \\
0 & 0 & 0 & 0 & 0 & 1 
\end{pmatrix},  \label{PCP_diffution_coeffiecents}
\end{equation}
\begin{equation} 
J =
\begin{pmatrix} 
 - \alpha b^{\dagger *} \mathcal{K}(c^{*}) & - \alpha a^{*} \mathcal{K}(c^{*}) & \phi_{1} & 0 & 0 & c^{*}\mathcal{V}'(c^{\dagger *}) \\
 - \alpha b^{\dagger *} \mathcal{K}(c^{*}) & - \alpha a^{*} \mathcal{K}(c^{*}) & \phi_{1} & 0 & 0 & c^{*}\mathcal{V}'(c^{\dagger *}) \\
  \alpha b^{\dagger *} \mathcal{K}(c^{*}) &  \alpha a^{*} \mathcal{K}(c^{*}) & - \phi_{1} & 0 & 0 & - c^{*}\mathcal{V}'(c^{\dagger *}) \\
0 & 0 & c^{\dagger*}\mathcal{V}'(c^{*}) & - \alpha b^{*} \mathcal{K}(c^{\dagger*}) & - \alpha a^{\dagger*} \mathcal{K}(c^{\dagger*}) & \phi_{2} \\
0 & 0 & c^{\dagger*}\mathcal{V}'(c^{*}) & - \alpha b^{*} \mathcal{K}(c^{\dagger*}) & - \alpha a^{\dagger*} \mathcal{K}(c^{\dagger*})  & \phi_{2} \\
0 & 0 & -c^{\dagger*}\mathcal{V}'(c^{*}) & \alpha b^{*} \mathcal{K}(c^{\dagger*}) &  \alpha a^{\dagger*} \mathcal{K}(c^{\dagger*})  & - \phi_{2}
\end{pmatrix}, \label{PCP_coefficent_matrix}
\end{equation}
and $\mathbf{U} = (\tilde{a}, \tilde{b}^{\dagger}, \tilde{c}, \tilde{a}^{\dagger}, \tilde{b}, \tilde{c}^{\dagger} )^{\intercal}$,
where $\phi_{1} := \mathcal{V}(c^{\dagger*}) - \alpha a^{*}b^{\dagger *} \mathcal{K}' (c^{*})$, and $\phi_{2} := \mathcal{V}(c^{*}) - \alpha a^{\dagger*}b^{*} \mathcal{K}' (c^{\dagger*})$.
 
Having assembled the Jacobian $J$ and diffusion matrix $D$, we now solve the eigenvalue problem associated with~\eqref{PCP_eq:linearised_system} mode by mode to obtain the dispersion relation. 
\label{PCP_dispersion}
Since the boundary conditions are of zero-flux type, we expand each perturbation in the cosine eigenfunctions of $\partial_{xx}$ on the (dimensionless) domain $(0,\ell)$,
\begin{equation}
\big(\tilde{a},\tilde{b}^{\dagger},\tilde{c},\tilde{a}^{\dagger},\tilde{b},\tilde{c}^{\dagger}\big)(x,t)
= \sum_{p\ge 0} \hat{\mathbf{U}}_{p}\, e^{\sigma_{p} t} \cos(q_{p} x),
\qquad q_{p} = \frac{p\pi}{\ell}, \quad p = 0,1,2,\dots,
\label{PCP_modal}
\end{equation}
so that \eqref{PCP_eq:linearised_system} reduces, mode by mode, to the algebraic eigenvalue problem $\sigma\,\hat{\mathbf{U}} = \big(J - q^{2} D\big)\hat{\mathbf{U}}$. 
The SUSS is linearly unstable if $\mathrm{Re}\,\sigma(q_{p}^{2}) > 0$ for some $p\ge 1$. 
The four conservation laws \eqref{a:cl}--\eqref{bdagger:cl} render $J$ rank-deficient: at $q=0$ it has a four-fold zero eigenvalue, so the homogeneous mode is neutrally stable and any instability is carried by a mode $q_{p}>0$.
 
\paragraph{Decoupled case ($\mathcal{V}'\equiv 0$).} 
We treat first the decoupled case, in which the unbinding rate $\mathcal{V}$ is a constant (so that $\mathcal{V}'\equiv 0$) and the turnover of each complex is independent of the opposite complex. 
This assumption is retained throughout the present section and through the single-punctum and array analyses of Sections~\ref{sec:single} and~\ref{sec:arrays}; the coupled case $\mathcal{V}'>0$, which generates sorting, is taken up in Section~\ref{PCP_polarity}.
With $\mathcal{V}$ constant the cross-terms in \eqref{PCP_coefficent_matrix} vanish and $J$ is block-diagonal, splitting the system into two independent triplets (each a closed set of three fields -- a monomer pair and their shared complex) $(\tilde{a},\tilde{b}^{\dagger},\tilde{c})$ and $(\tilde{a}^{\dagger},\tilde{b},\tilde{c}^{\dagger})$. 
Writing
\begin{equation}
\kappa_{1} := \alpha b^{\dagger *}\mathcal{K}(c^{*}), \qquad
\kappa_{2} := \alpha a^{*}\mathcal{K}(c^{*}), \qquad
\phi_{1} = \mathcal{V}(c^{\dagger *}) - \alpha a^{*} b^{\dagger *}\mathcal{K}'(c^{*}),
\label{PCP_block_entries}
\end{equation}
the unprimed block is $J_{1}-q^{2}D_{1}$ with
\begin{equation}
J_{1} =
\begin{pmatrix}
-\kappa_{1} & -\kappa_{2} & \phi_{1} \\
-\kappa_{1} & -\kappa_{2} & \phi_{1} \\
\phantom{-}\kappa_{1} & \phantom{-}\kappa_{2} & -\phi_{1}
\end{pmatrix},
\qquad
D_{1} = \mathrm{diag}(D_{a}, D_{b}, 1),
\label{PCP_block_matrix}
\end{equation}
and identically for the primed block with $a^{*}\!\to a^{\dagger *}$, $b^{\dagger *}\!\to b^{*}$, $c^{*}\!\to c^{\dagger *}$. 
The characteristic polynomial $\det(J_{1}-q^{2}D_{1}-\sigma I)=0$ is
\begin{equation}
\sigma^{3} + c_{2}(q^{2})\,\sigma^{2} + c_{1}(q^{2})\,\sigma + c_{0}(q^{2}) = 0,
\label{PCP_cubic}
\end{equation}
with coefficients
\begin{align}
c_{2} &= (\kappa_{1}+\kappa_{2}+\phi_{1}) + (D_{a}+D_{b}+1)\,q^{2}, \label{PCP_c2}\\
c_{1} &= (D_{a}D_{b}+D_{a}+D_{b})\,q^{4}
   + \big[D_{a}(\kappa_{2}+\phi_{1}) + D_{b}(\kappa_{1}+\phi_{1}) + (\kappa_{1}+\kappa_{2})\big]\,q^{2}, \label{PCP_c1}\\
c_{0} &= D_{a}D_{b}\,q^{4}\left(q^{2} + \frac{\kappa_{1}}{D_{a}} + \frac{\kappa_{2}}{D_{b}} + \phi_{1}\right). \label{PCP_c0}
\end{align}
The factor $q^{4}$ in \eqref{PCP_c0} reflects the two conserved masses of the triplet.
At $q=0$ the cubic~\eqref{PCP_cubic} reduces to $\sigma^{2}\big(\sigma + \kappa_{1}+\kappa_{2}+\phi_{1}\big)=0$, so the spatially homogeneous (well-mixed) state is stable iff
\begin{equation}
\kappa_{1}+\kappa_{2}+\phi_{1} > 0.
\label{PCP_wellmixed}
\end{equation}
For $q > 0$ the Routh--Hurwitz criterion~\citep{murray2001mathematical} gives stability iff $c_{2} > 0$, $c_{0}>0$ and $c_{2}c_{1}>c_{0}$. The first holds automatically whenever the well-mixed condition~\eqref{PCP_wellmixed} does. 
A real eigenvalue crosses zero precisely when $c_{0}=0$; as $c_{0}$ is a strictly increasing function of $q^{2}$, the SUSS is unstable to a stationary, \emph{mass-redistribution} mode -- a spatially non-uniform rearrangement of the conserved total density between regions of the junction, at fixed total mass, rather than a change in the total amount of bound or unbound material -- over the long-wavelength band
\begin{equation}
0 < q^{2} < q_{c}^{2}, \qquad
q_{c}^{2} = -\left(\phi_{1} + \frac{\kappa_{1}}{D_{a}} + \frac{\kappa_{2}}{D_{b}}\right),
\label{PCP_band}
\end{equation}
which is non-empty iff $\phi_{1} + \kappa_{1}/D_{a} + \kappa_{2}/D_{b} < 0$. 
Combining this with the well-mixed stability condition \eqref{PCP_wellmixed}, a diffusion-driven instability of an otherwise stable steady state requires
\begin{equation}
\frac{\kappa_{1}}{D_{a}} + \frac{\kappa_{2}}{D_{b}}
\;<\; -\phi_{1} \;<\; \kappa_{1} + \kappa_{2}.
\label{PCP_window}
\end{equation}
The window \eqref{PCP_window} is non-empty iff $\kappa_{1}/D_{a}+\kappa_{2}/D_{b}<\kappa_{1}+\kappa_{2}$, i.e.\ whenever the monomers diffuse faster than the complex ($D_{a},D_{b}>1$), and it widens as that disparity grows. 
The clustering feedback enters through $\phi_{1}$, the net linearised turnover of the complex (the constant unbinding rate $\mathcal{V}$ offset by the gain $\alpha a^{*}b^{\dagger *}\mathcal{K}'(c^{*})$ from concentration-dependent binding): a sufficiently positive $\mathcal{K}'(c^{*})$ drives $\phi_{1}$ negative and opens the band. 
The instability is therefore of long-wavelength (type-II) form~\citep{crosshohenberg1993}, with onset at $q\to 0$, rather than at a finite critical wavenumber as in a classical Turing system; in the language of \citet{brauns2020phase} it is a \emph{mass-redistribution instability}.
 
\paragraph{Symmetric steady state and the antisymmetric (sorting) mode.} 
The decoupled analysis above describes how puncta form, but says nothing about the \emph{relative} arrangement of the two complex orientations, since the two triplets are then independent. 
To see when the orientations interact -- the linear origin of sorting -- we now restore the coupling $\mathcal{V}'\neq 0$ at the symmetric steady state.
When $a_{T}=a^{\dagger}_{T}$ and $b_{T}=b^{\dagger}_{T}$ the system is invariant under the $\mathbb{Z}_{2}$ exchange $(a,b^{\dagger},c)\leftrightarrow(a^{\dagger},b,c^{\dagger})$, and the SUSS is symmetric, $c^{*}=c^{\dagger *}=:\bar{c}$, $a^{*}=a^{\dagger *}$, $b^{*}=b^{\dagger *}$. 
For $\mathcal{V}'\neq 0$ the two triplets are coupled only through the symmetric pair of off-diagonal entries in \eqref{PCP_coefficent_matrix}, so $J-q^{2}D$ block-diagonalises in the parity basis $\hat{\mathbf{U}}_{\pm}=\tfrac{1}{\sqrt{2}}\big[(\tilde{a},\tilde{b}^{\dagger},\tilde{c})\pm(\tilde{a}^{\dagger},\tilde{b},\tilde{c}^{\dagger})\big]$ into two blocks of the form \eqref{PCP_block_matrix}, with $\phi_{1}$ replaced by
\begin{equation}
\phi_{\pm} = \phi \pm \omega, \qquad
\phi = \mathcal{V}(\bar{c}) - \alpha a^{*}b^{\dagger *}\mathcal{K}'(\bar{c}), \qquad
\omega = \bar{c}\,\mathcal{V}'(\bar{c}).
\label{PCP_parity}
\end{equation}
Each parity block has the cubic \eqref{PCP_cubic}--\eqref{PCP_c0} with $\phi_{1}\to\phi_{\pm}$, and hence an instability band \eqref{PCP_band} with critical wavenumber
$q_{c,\pm}^{2}=-\big(\phi_{\pm}+\kappa_{1}/D_{a}+\kappa_{2}/D_{b}\big)$. 
The symmetric ($+$) mode $\tilde{c}+\tilde{c}^{\dagger}$ describes in-phase clustering of the two complexes (co-localised puncta); the antisymmetric ($-$) mode $\tilde{c}-\tilde{c}^{\dagger}$ describes their segregation to opposite regions of the same contact (sorting). 
We emphasise that this is segregation \emph{within} one cell--cell contact, and not the cell-scale asymmetry between opposite edges of a cell. 
Since $\phi_{-}<\phi_{+}$ iff $\omega>0$, the antisymmetric (sorting) mode is the first to lose stability precisely when $\mathcal{V}'(\bar{c})>0$, i.e.\ when each complex promotes the unbinding of the complex of opposite orientation -- the competitive, destabilising cross-interaction posited for Prickle in Section~\ref{sec:introduction}. 
Conversely $\mathcal{V}'<0$ favours the co-localised mode, and the decoupled choice $\mathcal{V}'\equiv 0$ makes the two parities degenerate, so at linear order the model does not distinguish sorting from co-localisation -- consistent with the simultaneous clustering of $c$ and $c^{\dagger}$ seen in Figure~\ref{AB_onset_numerics}.

\begin{figure}[htbp]
\centering
\includegraphics[width=0.8\textwidth]{ 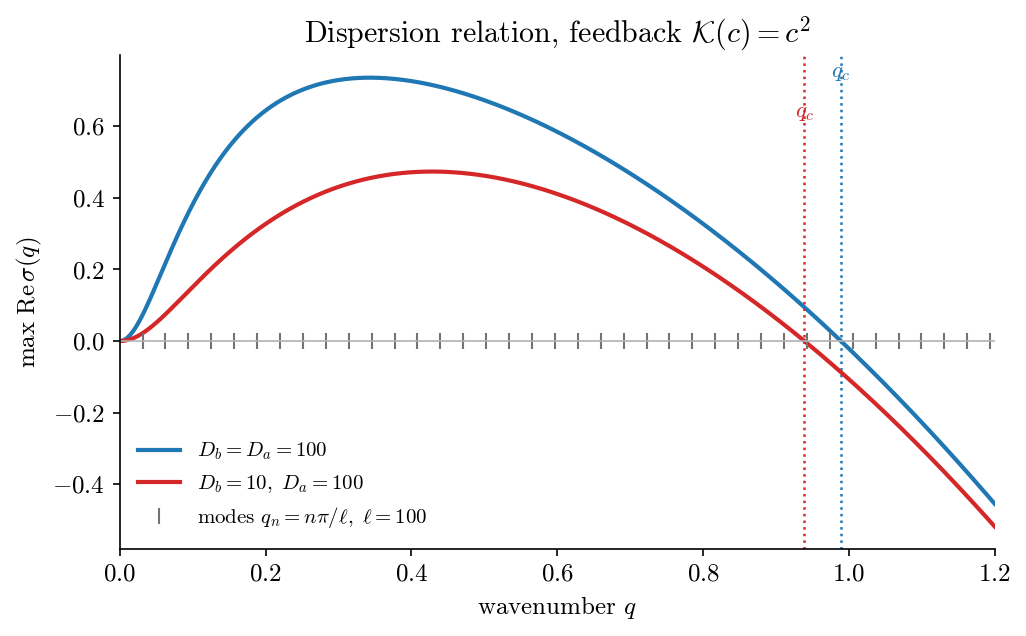}
\caption{\textbf{Dispersion relation of the decoupled block.} 
Largest real part of the growth rate $\sigma$ obtained from the cubic \eqref{PCP_cubic}--\eqref{PCP_c0}, for the feedback $\mathcal{K}(c)=c^{2}$, $\mathcal{V}=1$, with representative $\alpha=0.14$ and symmetric totals $a_{T}=b^{\dagger}_{T}=3.9$ (so that $c^{*}\approx 2.02$, $\kappa_{1}=\kappa_{2}\approx 1.07$ and $\phi_{1}=-1$). 
The two curves correspond to $D_{a}=D_{b}=100$ (cf.\ Figure~\ref{AB_onset_numerics}(a)) and $D_{a}=100$, $D_{b}=10$ (cf.\ Figure~\ref{AB_onset_numerics}(b)). 
Vertical dotted lines mark the band edge $q_{c}$ of \eqref{PCP_band}; the tick marks show the admissible modes $q_{p}=p\pi/\ell$ for $\ell=100$. 
The instability occupies a long-wavelength band $0<q<q_{c}$.}
\label{AB_dispersion}
\end{figure}
In summary, the uniform state is destabilised by a stationary, long-wavelength band of modes whenever the monomers diffuse sufficiently faster than the complex and the clustering feedback is strong enough to drive $\phi_{1}$ negative, with the antisymmetric (sorting) mode leading the instability once $\mathcal{V}'>0$. 
The next subsection interprets this onset condition physically, through the well-mixed bistable kinetics on the mass-conservation manifold, before we turn to the nonlinear structure of the resulting puncta in Section~\ref{sec:single}.

\subsection{Well-mixed kinetics and bistability} \label{PCP_bistability}
 
\begin{figure}[htbp]
\centering
\includegraphics[width=\textwidth]{ 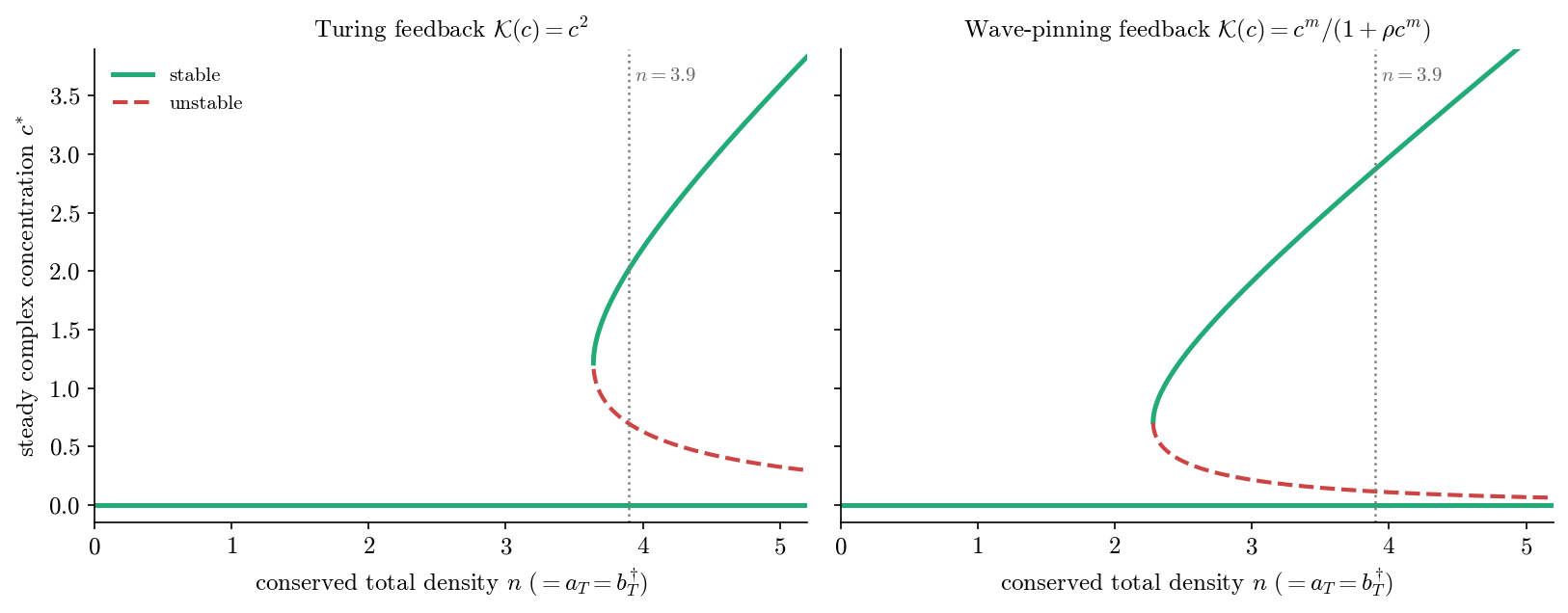}
\caption{\textbf{Reactive nullclines and bistability.} Steady complex concentration $c^{*}$ as a function of the conserved total density $n=a_{T}=b^{\dagger}_{T}$, from \eqref{PCP_nullcline}, with stable (solid) and unstable (dashed) branches; the empty state $c=0$ is a stable equilibrium in both cases. 
\emph{Left:} feedback $\mathcal{K}(c)=c^{2}$ ($\alpha=0.14$), fold at $n_{\mathrm{f}}\approx 3.6$. 
\emph{Right:} saturating feedback $\mathcal{K}(c)=c^{m}/(1+\rho c^{m})$, $m=2$, $\rho=0.1$ ($\alpha=0.6$), fold at $n_{\mathrm{f}}\approx 2.3$. 
The dotted line marks the operating density $n=3.9$. 
The values of $\alpha$ are representative; Figure~\ref{AB_onset_numerics} uses $\alpha=0.14$ for the unbounded feedback and $\alpha=0.6$ for the saturating one.}
\label{AB_nullcline}
\end{figure}

The instability \eqref{PCP_band} is most naturally interpreted through the spatially homogeneous (well-mixed) kinetics on the mass-conservation manifold. 
Recall from Section~\ref{PCP_dispersion} that the decoupled system splits into two triplets, and that the primed triplet $(a^{\dagger},b,c^{\dagger})$ behaves identically to the unprimed one; we therefore work with the unprimed fields $(a,b^{\dagger},c)$ throughout. 
Within that triplet the conservation laws fix $a = a_{T} - c$ and $b^{\dagger} = b^{\dagger}_{T} - c$, so the reaction dynamics reduce to the scalar equation
\begin{equation}
\dot{c} = F(c) := \alpha\,\mathcal{K}(c)\,(a_{T} - c)(b^{\dagger}_{T} - c)
   - \mathcal{V}(c^{\dagger})\,c .
\label{PCP_scalar}
\end{equation}
For the decoupled choice $\mathcal{V}\equiv 1$ (Section~\ref{sec:feedback}) -- the setting of the clustering and array simulations presented below in Figures~\ref{AB_onset_numerics} and~\ref{AB_spike}--\ref{AB_coarsening}; the sorting-generating coupling $\mathcal{V}'>0$ is deferred to Section~\ref{PCP_polarity} and Figure~\ref{AB_sorting} -- and symmetric totals $a_{T}=b^{\dagger}_{T}=n$, equation \eqref{PCP_scalar} is a bistable scalar kinetics
with the conserved total density $n$ as control parameter. 
Its equilibria solve $\alpha\mathcal{K}(c)(n-c)^{2}=c$, which we write as the \emph{reactive nullcline}
\begin{equation}
n = \mathcal{N}(c) := c + \sqrt{\frac{c}{\alpha\,\mathcal{K}(c)}},
\label{PCP_nullcline}
\end{equation}
on the physical branch $n>c$. Both feedback choices of Section~\ref{sec:feedback}, the unbounded \eqref{PCP_feedback_turing} and the saturating \eqref{PCP_feedback_sat}, satisfy $\mathcal{K}(0)=\mathcal{K}'(0)=0$, so the empty state $c=0$ is always an equilibrium and is stable ($F'(0)=-1<0$). 
A second, high-$c$ stable branch is created at a saddle-node (fold) where $\mathcal{N}'(c)=0$. 
The kinetics are therefore bistable for every $n$ above the fold density $n_{\mathrm{f}}=\min_{c}\mathcal{N}(c)$, with an unstable middle branch separating the two basins (Figure~\ref{AB_nullcline}).
 
For the unbounded feedback $\mathcal{K}(c)=c^{2}$ of \eqref{PCP_feedback_turing} the nullcline is $\mathcal{N}(c)=c+(\alpha c)^{-1/2}$, with a single fold at $c_{\mathrm{f}}=(2\sqrt{\alpha})^{-2/3}$. 
The upper-branch steady state then satisfies $\alpha c^{*}(n-c^{*})^{2}=1$, whence $\kappa_{1}=\kappa_{2}=c^{*}/(n-c^{*})$ and, independently of $n$ and $\alpha$,
\begin{equation}
\phi_{1} = 1 - 2\alpha c^{*}(n-c^{*})^{2} = -1 .
\label{PCP_phiminus1}
\end{equation}
The well-mixed state is stable precisely on this upper branch ($c^{*}>n/3$), and the band \eqref{PCP_band} reduces to $q_{c}^{2}=1-\dfrac{c^{*}}{\,n-c^{*}}\big(D_{a}^{-1}+D_{b}^{-1}\big)$, which is close to unity for the diffusivities of Figure~\ref{AB_onset_numerics}; a wide band of modes is therefore unstable (Figure~\ref{AB_dispersion}). 
The saturating feedback $\mathcal{K}(c)=c^{m}/(1+\rho c^{m})$ of \eqref{PCP_feedback_sat} has the same qualitative structure but, through saturation, lowers the fold density $n_{\mathrm{f}}$ and widens the bistable range, so that at the operating point $n=3.9$ the system sits well above threshold (Figure~\ref{AB_nullcline}, right).
 
The combination of bistable kinetics, mass conservation, and faster diffusion of the monomers than of the complex ($D_{a},D_{b}>1$) is exactly the set of ingredients identified by \citet{mori2008wave} for \emph{wave pinning}. 
This supplies the real-space content of the long-wavelength mass-redistribution instability identified in Section~\ref{PCP_dispersion}: regions of locally elevated total density settle onto the upper branch of $\mathcal{N}(c)$ while depleted regions relax towards $c\approx 0$, and diffusion of the fast monomers transports mass between them until a stationary, pinned front forms. 
Both feedbacks share this mechanism and differ only in the position of the fold, and hence in how far the operating density lies above threshold. 
The contrast between the spike solutions of Figure~\ref{AB_onset_numerics}(a) and the mesa solutions of Figure~\ref{AB_onset_numerics}(c) (both obtained by integrating the full six-species system from a near-uniform state at equal monomer diffusivities) is therefore one of degree rather than of mechanism: a broad band of unstable modes seeds many interfaces, which are expected to coarsen towards a single front on sufficiently large domains and long times. 
A quantitative prediction of the pinned-front position, via the sharp-interface (equal-area, or Maxwell) construction standard for bistable fronts~\citep{mori2008wave, mckay2012stability} in the limit $D_{a},D_{b}\gg \ell^{2}$, is carried out in Section~\ref{PCP_pinned}.

\begin{figure}[htbp]
\centering
\includegraphics[width=\textwidth]{ 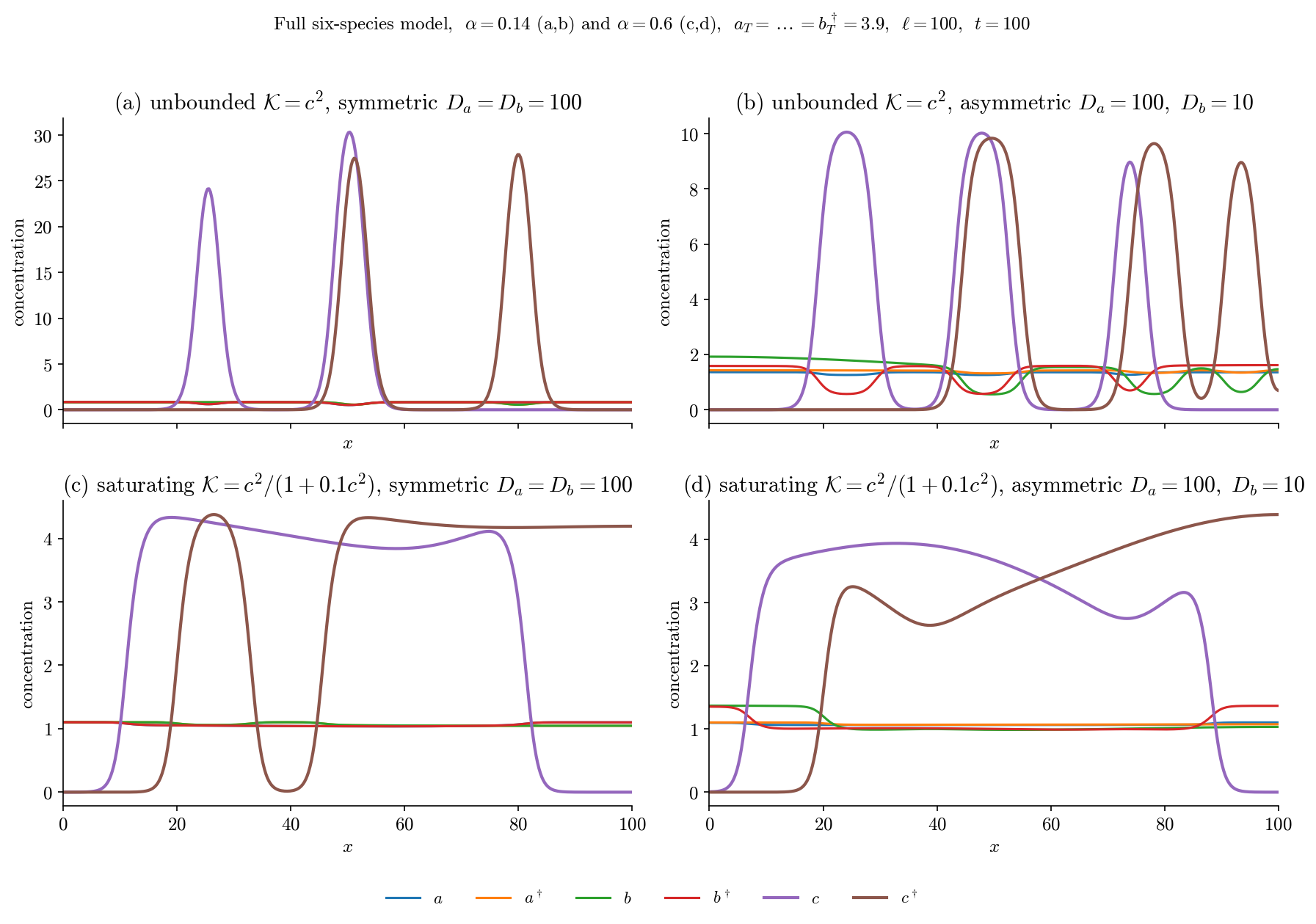}
\caption{\textbf{Numerical solution of the full six-species model \eqref{a**:eq}--\eqref{cdagger**:eq} under the two clustering feedbacks.}
Integration from a near-uniform state on $\ell=100$ with symmetric totals $a_{T}=a^{\dagger}_{T}=b_{T}=b^{\dagger}_{T}=3.9$, $\mathcal{V}=1$, perturbation amplitude $\epsilon=2.5$, shown at $t=100$; masses conserved to machine precision.
\emph{Top row}, the unbounded (Turing) feedback $\mathcal{K}(c)=c^{2}$ with $\alpha=0.14$: (a) symmetric diffusion $D_{a}=D_{b}=100$, where the complexes form narrow \emph{spikes} (puncta, Sections~\ref{PCP_homodimer}--\ref{PCP_spike}) on a depleted monomer background; (b) asymmetric diffusion $D_{a}=100$, $D_{b}=10$, where the less mobile monomer is depleted only locally, the spikes are shorter and more numerous, and the two monomers have distinct screening lengths (Section~\ref{PCP_asym}).
\emph{Bottom row}, the saturating (wave-pinning) feedback $\mathcal{K}(c)=c^{m}/(1+\rho c^{m})$ with $m=2$, $\rho=0.1$, $\alpha=0.6$: (c) symmetric and (d) asymmetric diffusion at the same diffusivities, where the complexes instead form broad \emph{mesas} -- pinned high-density domains separated by empty regions (Section~\ref{PCP_pinned}) -- the same mass-redistribution instability selecting a different morphology under bounded feedback. The plateau approaches the Maxwell value $c^{*}_{+}\approx4.79$ as $D_{a},D_{b}$ increase.
The several coexisting spikes in (a) and (b) are the metastable, partially coarsened states of Section~\ref{PCP_competition}, shown here at a finite time: every like-oriented array is linearly unstable, so on this domain the true steady state of each triplet carries a single spike, and the configurations shown persist only because coarsening is slow at these diffusivities. 
Since $\mathcal{V}'=0$ the $c$ and $c^{\dagger}$ fields evolve independently, so their relative placement reflects the initial perturbation, not sorting (which requires $\mathcal{V}'>0$, Section~\ref{PCP_dispersion}).}
\label{AB_onset_numerics}
\end{figure}

\section{The structure of a single punctum} \label{sec:single}

Section~\ref{sec:onset} established \emph{when} the uniform state gives way to patterning, but not \emph{what} the resulting localised structures look like. 
We now turn to that nonlinear question, analysing the profile of a single punctum in the decoupled system ($\mathcal{V}\equiv 1$). 
The well-mixed analysis of Section~\ref{PCP_bistability} already anticipates two regimes: a saturating feedback admits two stable homogeneous states and so supports a pinned front between them, whereas an unbounded feedback has only the empty state as a stable rest point and so cannot pin a front. 
We treat these in turn: the broad, pinned \emph{mesa} of the saturating feedback (Section~\ref{PCP_pinned}) and the narrow, mass-limited \emph{spike} of the unbounded feedback (Sections~\ref{PCP_homodimer}--\ref{PCP_spike}). 
For the spike we show that the profile cleanly separates a punctum's amplitude, width and spacing. 
Throughout we exploit the fast-monomer regime, in which the monomers are nearly uniform and the analysis becomes tractable.

\subsection{Pinned fronts and mesas} \label{PCP_pinned}
 
\begin{figure}[htbp]
\centering
\includegraphics[width=0.72\textwidth]{ 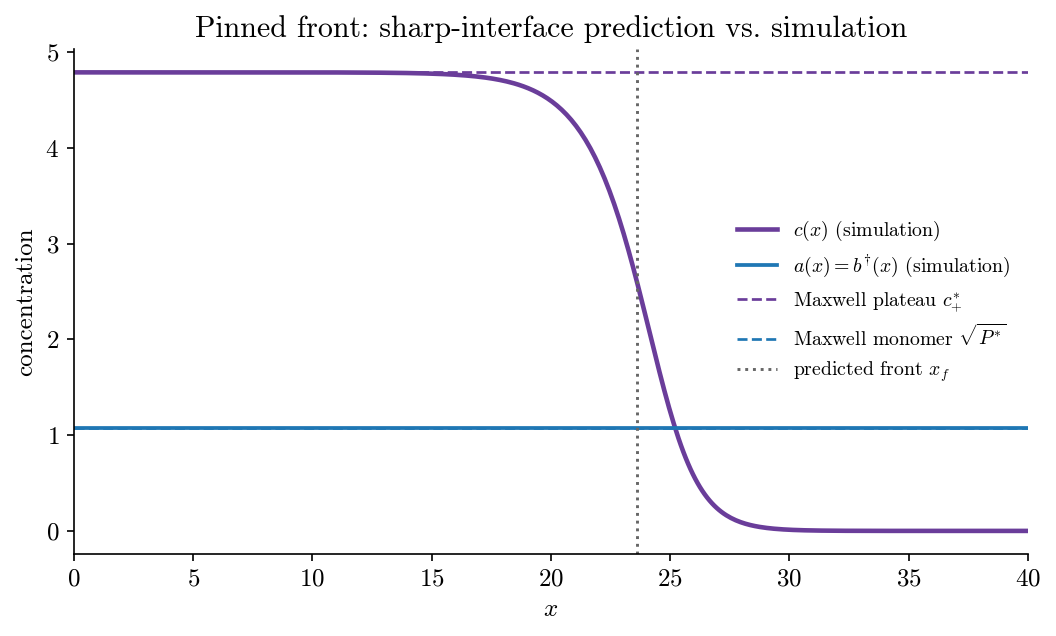}
\caption{\textbf{Pinned front: sharp-interface theory versus simulation.} 
Steady profiles of the complex $c(x)$ (purple) and the monomers $a(x)=b^{\dagger}(x)$ (blue) from a direct simulation of the decoupled triplet with the saturating feedback ($m=2$, $\rho=0.1$, $\alpha=0.6$) at $D_{a}=D_{b}=10^{5}$ on a domain $\ell=40$ with symmetric total density $n=3.9$. 
Dashed lines: the Maxwell plateau $c_{+}^{*}$ and monomer level $\sqrt{P^{*}}$ from \eqref{PCP_maxwell}; dotted line: the front position $x_{f}$ from \eqref{PCP_position}. 
The simulated values ($c_{+}=4.79$, $\bar{a}=1.07$, $x_{f}/\ell=0.59$) match the predictions to within a fraction of a percent.}
\label{AB_pinning_maxwell}
\end{figure}

When the monomers diffuse much faster than the complex relative to the domain, $D_{a},D_{b}\gg\ell^{2}$, they equilibrate across the whole domain and become uniform to leading order, $a\to\bar{a}$, $b^{\dagger}\to\bar{b}$. 
This regime is the relevant one for planar polarity puncta: the clustered complexes are the slowly diffusing, low-turnover species, whereas the unbound monomers explore the junction rapidly, so on the scale of a single cell--cell contact the monomer pools are well mixed while the complex remains spatially structured. 
Working to leading order in the resulting separation of scales isolates the complex dynamics, with the monomers entering only through their (slowly varying) mean levels $\bar{a},\bar{b}$.
The complex then obeys the scalar bistable equation
\begin{equation}
c_{t} = c_{xx} + g(c), \qquad g(c) := \alpha\,\mathcal{K}(c)\,\bar{a}\,\bar{b} - c,
\label{PCP_scalar_front}
\end{equation}
with $\bar{a}\bar{b}$ a slowly varying parameter set by the conserved masses. 
A pinned interface requires $g$ to possess two stable zeros: this holds for the saturating feedback $\mathcal{K}(c)=c^{m}/(1+\rho c^{m})$ (zeros $c_{-}=0$, an unstable $c_{\mathrm{m}}$ and a stable $c_{+}$), but \emph{not} for $\mathcal{K}(c)=c^{2}$, whose upper zero is unstable -- the high state is then mass-limited rather than pinned, giving the spikes of Section~\ref{PCP_homodimer}. 
The stationary front of \eqref{PCP_scalar_front} obeys the equal-area (Maxwell) condition $\int_{0}^{c_{+}}g\,\mathrm{d}c=0$, the standard sharp-interface selection criterion for a front between two stable states of a bistable reaction--diffusion equation~\citep{mori2008wave, mckay2012stability}; here, with $g(c_{+})=0$, it fixes the plateau height $c_{+}^{*}$ and the monomer product $P^{*}:=\bar{a}\bar{b}$,
\begin{equation}
\int_{0}^{c_{+}^{*}}\mathcal{K}(c)\,\mathrm{d}c = \tfrac{1}{2}\,c_{+}^{*}\,\mathcal{K}(c_{+}^{*}),
\qquad
P^{*} = \frac{c_{+}^{*}}{\alpha\,\mathcal{K}(c_{+}^{*})} ,
\label{PCP_maxwell}
\end{equation}
the first relation independent of $\alpha$ and of the masses; for $m=2$, $\rho=0.1$, $\alpha=0.6$ it gives $c_{+}^{*}\approx 4.79$ and $\sqrt{P^{*}}\approx 1.07$. 
Mass conservation then fixes the location of the front. 
Let $x_{f}$ denote the front position, so that the mesa occupies the high plateau $c\approx c_{+}^{*}$ on $0<x<x_{f}$ and the empty state $c\approx 0$ on $x_{f}<x<\ell$ (Figure~\ref{AB_pinning_maxwell}). 
In the fast-monomer limit the monomers are uniform, $\bar{a}=\bar{b}=\sqrt{P^{*}}$, so evaluating the conserved total~\eqref{a:cl} on this step profile gives
\begin{equation}
n = \frac{1}{\ell}\int_{0}^{\ell}\!\big(a+c\big)\,\mathrm{d}x
\;\approx\; \sqrt{P^{*}} + c_{+}^{*}\,\frac{x_{f}}{\ell},
\label{PCP_position_balance}
\end{equation}
since the uniform monomer contributes $\bar{a}=\sqrt{P^{*}}$ and the plateau contributes $c_{+}^{*}$ over a fraction $x_{f}/\ell$ of the domain. 
Solving for the front position,
\begin{equation}
\frac{x_{f}}{\ell} = \frac{n-\sqrt{P^{*}}}{c_{+}^{*}},
\label{PCP_position}
\end{equation}
so the punctate fraction of the junction grows linearly with the available mass, for $\sqrt{P^{*}}<n<\sqrt{P^{*}}+c_{+}^{*}$. 
Figure~\ref{AB_pinning_maxwell} confirms \eqref{PCP_maxwell}--\eqref{PCP_position} against a direct simulation at $D_{a}=D_{b}=10^{5}$, $\ell=40$, to within a fraction of a percent. 
This single-front regime requires $D_{a},D_{b}\gg\ell^{2}$; the simulations of Figure~\ref{AB_onset_numerics}, with $D_{a}$ comparable to $\ell$, instead show several coexisting interfaces that coarsen on longer timescales. 
The mesa is thus the localised state selected by the \emph{saturating} feedback: it represents a broad punctum whose protein density is clamped at the fixed plateau value $c_{+}^{*}$, while the region it covers simply expands as more complex becomes available (equation~\eqref{PCP_position}). 
When the clustering feedback instead grows without bound, no such plateau can be sustained -- there is no stable high-density state for a front to pin against -- and the complex piles up locally into a narrow, tall peak whose height is capped only by the finite supply of protein. 
We identify this peak with a single, sharply defined punctum. 
We turn to it next. Section~\ref{PCP_homodimer} begins with the simplest single-reservoir caricature, which strips the mechanism down to its essentials before we reinstate the full heterodimer.

\subsection{A single-reservoir warm-up} \label{PCP_homodimer}
 
Clustering into puncta is not special to planar polarity: homophilic adhesion receptors such as E-cadherin also concentrate into discrete puncta at cell--cell junctions~\citep{quang2013principles}. 
The simplest caricature of such a punctum replaces the heterodimer reactions of Section~\ref{PCP_model} by a single self-associating species forming a \emph{trans}-homodimer, $A+A^{\dagger}\rightleftharpoons C$, with $A$ and $A^{\dagger}$ the same molecule presented on the two faces of the junction. 
With equal totals the faces are equivalent, $a=a^{\dagger}$, and the dynamics reduce to one monomer reservoir coupled to the complex,
\begin{equation}
a_{t}=D\,a_{xx}-\alpha\,\mathcal{K}(c)\,a^{2}+\mathcal{V}c,\qquad
c_{t}=c_{xx}+\alpha\,\mathcal{K}(c)\,a^{2}-\mathcal{V}c,
\label{PCP_AA_rd}
\end{equation}
conserving the single mass $a+c$. 
This is precisely the single-reservoir model to which the symmetric heterodimer collapses when $D_{a}=D_{b}$, and it isolates the spike mechanism in its simplest form. 
We therefore analyse it on its own. 
Stripped of the second reservoir, it carries all the algebra of the spike profile -- height, width and existence threshold -- in its cleanest form, and the heterodimer results of Sections~\ref{PCP_spike}--\ref{PCP_asym} read most transparently as additions to this baseline. 
We carry out that analysis here and identify, by contrast, exactly what the second reservoir contributes.
 
For the unbounded feedback $\mathcal{K}(c)=c^{2}$ with $\mathcal{V}=1$ and fast monomer diffusion the inner problem is $c_{xx}-c+\alpha\bar{a}^{2}c^{2}=0$; the rescaling $c=(\alpha\bar{a}^{2})^{-1}w$ reduces this to the canonical spike equation $w''-w+w^{2}=0$, whose localised (homoclinic) solution is $w=\tfrac{3}{2}\operatorname{sech}^{2}(y/2)$~\citep{iron2001stability}. 
We stress that nothing in this reduction is new: it is the standard activator--substrate spike, and we rehearse it only to fix the baseline against which the two-reservoir effects of Sections~\ref{PCP_spike}--\ref{PCP_asym} are measured. 
Thus
\begin{equation}
c_{\max}=\frac{3}{2\alpha\bar{a}^{2}},\qquad n=\bar{a}+\frac{6}{\alpha\bar{a}^{2}\ell},
\label{PCP_AA_spike}
\end{equation}
where $n=a_{T}=\tfrac{1}{\ell}\int_{0}^{\ell}(a+c)\,\mathrm{d}x$ is the single conserved total density of the homodimer; mass conservation fixes it as the sum of the depleted reservoir level $\bar{a}$ and the contribution $6/(\alpha\bar{a}^{2}\ell)$ of the spike (computed exactly as in Section~\ref{PCP_spike} below). 
The height is now set by a \emph{single} reservoir, $c_{\max}\propto\bar{a}^{-2}$. 
The right-hand side of \eqref{PCP_AA_spike} has a minimum at $\bar{a}_{*}=(12/\alpha\ell)^{1/3}$, so a spike exists only above the threshold mass $n_{\mathrm{f}}=\tfrac{3}{2}(12/\alpha\ell)^{1/3}$. 
This is an \emph{existence fold}: a saddle-node bifurcation in the spike amplitude as a function of the control parameter $n$~\citep{iron2001stability, mckay2012stability}, at which a tall spike on a depleted reservoir and a short spike on a fuller reservoir are created together. 
It coincides with the symmetric heterodimer threshold derived below in \eqref{PCP_spike_fold}, since the two problems share the same scalar mass balance when $\bar{a}=\bar{b}$. 
The same competition analysis we develop below for the heterodimer (Section~\ref{PCP_competition}) applies here, now with a single substrate Green's function; as shown there, any $K$-spike array ($K\geq2$) is linearly unstable and coarsens towards a single punctum at the parameter-free rate $\lambda_{\mathrm{comp}}=\alpha\bar{a}^{3}D/(3\ell)$, set by the lone screening length $\sqrt{D}$.
Because the species is homophilic and symmetric, the homodimer has no analogue of the antisymmetric (sorting) mode of Section~\ref{PCP_dispersion} -- it can cluster, but it cannot sort. 
The symmetric simulations of Figures~\ref{AB_spike} and~\ref{AB_competition} ($D_{a}=D_{b}$, hence $a=b^{\dagger}$) are simultaneously simulations of \eqref{PCP_AA_rd}: a direct integration reproduces $c_{\max}=3/(2\alpha\bar{a}^{2})$ ($19.7$ against $19.9$ predicted) and $\lambda_{\mathrm{comp}}=\alpha\bar{a}^{3}D/(3\ell)$ to within about ten percent in the shadow regime.
 
The heterodimer of the following sections keeps this spike skeleton but adds the two ingredients that make planar polarity distinctive: a \emph{second}, independently conserved reservoir (Sections~\ref{PCP_spike}--\ref{PCP_asym}), and -- once $\mathcal{V}'\neq0$ -- the antisymmetric mode by which the two complex species segregate to opposite regions of the contact.
With the mechanism fixed by this warm-up, we now restore the two reservoirs and construct the heterodimer spike in full, reading off how its amplitude, width and spacing depend on the model parameters.
 
\subsection{Mass-limited spikes: a model for individual puncta} \label{PCP_spike}
 
The pinned front of Section~\ref{PCP_pinned} describes a broad domain of elevated complex density. 
For the unbounded feedback $\mathcal{K}(c)=c^{2}$, by contrast, the localised state is a narrow \emph{spike}, which we identify with an individual punctum. 
Whereas the mesa is a heteroclinic connection between two stable states, the spike is a homoclinic orbit returning to the empty state $c=0$; its construction follows the singular-perturbation analysis of spike patterns in activator--substrate systems \citep{gierer1972theory, iron2001stability}, adapted to the present mass-conserving (substrate-depletion) setting \citep{mckay2012stability}.
As in Section~\ref{PCP_pinned} we work in the regime of fast monomer diffusion, $D_{a},D_{b}\gg 1$, and consider the decoupled triplet \eqref{a**:eq},\,\eqref{bdagger**:eq},\,\eqref{c**:eq} with $\mathcal{K}(c)=c^{2}$ and $\mathcal{V}=1$.
We retain the decoupled choice $\mathcal{V}=1$ throughout this section because we are concerned here only with the formation and structure of a \emph{single} punctum, for which the two orientations evolve independently (Section~\ref{PCP_dispersion}). 
The cross-coupling $\mathcal{V}'>0$ leaves the single-spike profile unchanged -- a lone punctum sits where the opposite complex is essentially absent -- and acts only on the relative arrangement of two overlapping puncta; we therefore defer it to Section~\ref{PCP_polarity}, where it generates sorting.
 
\paragraph{Outer region.} 
Away from the spike the monomers are spatially uniform to leading order, $a\to\bar{a}$ and $b^{\dagger}\to\bar{b}$ (with corrections of order $D_{a}^{-1}$, $D_{b}^{-1}$ localised near the spike), while the complex relaxes to the empty state $c\approx 0$, which is an exact reactive equilibrium since $\mathcal{K}(0)=0$.
 
\begin{figure}[htbp]
\centering
\includegraphics[width=0.72\textwidth]{ 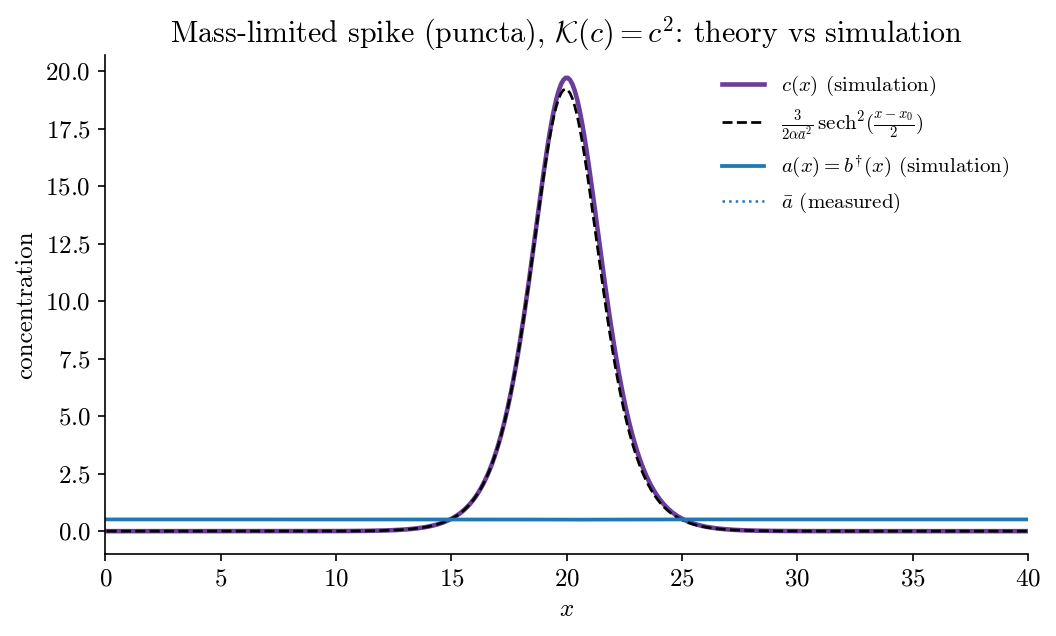}
\caption{\textbf{Mass-limited spike (punctum) for the unbounded feedback $\mathcal{K}(c)=c^{2}$.} 
Steady profile of the complex $c(x)$ (purple) from a simulation of the decoupled triplet with $\mathcal{V}=1$, $\alpha=0.3$, $D_{a}=D_{b}=2000$ on a domain $\ell=40$ and symmetric total density $n=2.5$; the dashed curve is the asymptotic profile \eqref{PCP_spike_profile} evaluated at the simulated reservoir $\bar{a}$, and the blue line is the depleted monomer reservoir. 
The measured peak ($c_{\max}\approx 19.7$) and full width at half maximum ($3.50$) match \eqref{PCP_spike_profile}, for which $\mathrm{FWHM}=4\, \mathrm{arccosh}\sqrt{2}\approx 3.53$. 
The reservoir relaxes to $\bar{a}\approx 0.51$, the tall branch of \eqref{PCP_spike_sym}.}
\label{AB_spike}
\end{figure}

\paragraph{Inner region.} 
Near the spike centre $x_{0}$ the complex varies on the $O(1)$ length set by its own diffusivity while the monomers remain at their reservoir values $\bar{a}$, $\bar{b}$, so the steady profile satisfies
\begin{equation}
c_{xx} - c + \alpha\,\bar{a}\,\bar{b}\,c^{2} = 0, \qquad c\to 0\ \text{as}\ |x-x_{0}|\to\infty,
\label{PCP_spike_inner}
\end{equation}
which is exactly the homodimer problem of Section~\ref{PCP_homodimer} with the product $\bar{a}\bar{b}$ in place of $\bar{a}^{2}$. 
The same scaling $c=(\alpha\bar{a}\bar{b})^{-1}w$, $w=\tfrac{3}{2}\operatorname{sech}^{2}(y/2)$, then gives
\begin{equation}
c(x) \sim \frac{3}{2\alpha\bar{a}\bar{b}}\,
\operatorname{sech}^{2}\!\left(\frac{x-x_{0}}{2}\right),
\qquad
c_{\max} = \frac{3}{2\alpha\bar{a}\bar{b}},
\label{PCP_spike_profile}
\end{equation}
so the punctum height is inversely proportional to the product of the two monomer reservoirs.
The symmetric reduction $\bar{a}=\bar{b}$ of this prediction is confirmed directly by the simulation of Figure~\ref{AB_spike} (Section~\ref{PCP_spike}), in which the measured peak matches $3/(2\alpha\bar{a}^{2})$ to within the finite-$D_{a}$ correction.
 
\paragraph{Mass conservation.} 
The reservoir levels $\bar{a}$, $\bar{b}$ are not free: they are fixed by the conserved masses \eqref{a:cl},\,\eqref{bdagger:cl}. Integrating the spike profile \eqref{PCP_spike_profile} and using $\int_{-\infty}^{\infty}\operatorname{sech}^{2}(x/2)\,\mathrm{d}x=4$, a single spike carries complex mass
\begin{equation*}
\int c\,\mathrm{d}x \sim \frac{3}{2\alpha\bar{a}\bar{b}}\int_{-\infty}^{\infty}
\operatorname{sech}^{2}\!\Big(\tfrac{x-x_{0}}{2}\Big)\,\mathrm{d}x
= 4\,c_{\max} = \frac{6}{\alpha\bar{a}\bar{b}}.
\end{equation*}
In the outer region the monomers are uniform at $\bar{a}$ and $\bar{b}$, while the complex is negligible away from the narrow spike. 
Evaluating the conserved total $a_{T}$ of \eqref{a:cl} therefore splits into a uniform monomer part $\tfrac{1}{\ell}\int a\,\mathrm{d}x\approx\bar{a}$ and the localised complex part $\tfrac{1}{\ell}\int c\,\mathrm{d}x\approx 6/(\alpha\bar{a}\bar{b}\,\ell)$, and likewise for $b^{\dagger}_{T}$ from \eqref{bdagger:cl} (recall that the complex $C$ consumes one $A$ and one $B^{\dagger}$, so $c$ enters both conservation laws), giving
\begin{equation}
a_{T} = \bar{a} + \frac{6}{\alpha\bar{a}\bar{b}\,\ell}, \qquad
b^{\dagger}_{T} = \bar{b} + \frac{6}{\alpha\bar{a}\bar{b}\,\ell}.
\label{PCP_spike_mass}
\end{equation}
These two relations determine $\bar{a}$ and $\bar{b}$ implicitly in terms of the prescribed totals.
Subtracting them, the spike contribution cancels and $\bar{a}-\bar{b}=a_{T}-b^{\dagger}_{T}$, so the reservoirs inherit the imbalance of the totals; for $K$ identical spikes the spike term is multiplied by $K$.
The significance is that the difference of the two reservoirs is pinned by the conserved totals alone, independently of how much complex has formed or of how many spikes there are: because each complex removes one $A$ and one $B^{\dagger}$ in equal measure, clustering cannot alter the \emph{difference} $\bar{a}-\bar{b}$, only the common level. 
An unequal supply of the two monomers ($a_{T}\neq b^{\dagger}_{T}$) is therefore felt as a permanent offset between the reservoirs, with the scarcer monomer held at the lower level -- the feature that, once $D_{a}\neq D_{b}$, gives the heterodimer its second screening length (Section~\ref{PCP_asym}).
 
\paragraph{Symmetric case and existence threshold.} When $a_{T}=b^{\dagger}_{T}=n$ we have $\bar{a}=\bar{b}$ and \eqref{PCP_spike_mass} reduces to
\begin{equation}
n = \bar{a} + \frac{6}{\alpha\bar{a}^{2}\,\ell}.
\label{PCP_spike_sym}
\end{equation}
The right-hand side has a minimum at $\bar{a}_{*}=(12/\alpha\ell)^{1/3}$, so a single spike exists only above the threshold mass
\begin{equation}
n > n_{\mathrm{f}} = \frac{3}{2}\left(\frac{12}{\alpha\ell}\right)^{1/3},
\label{PCP_spike_fold}
\end{equation}
and for $n>n_{\mathrm{f}}$ equation \eqref{PCP_spike_sym} has two solutions: a tall spike on a depleted reservoir ($\bar{a}<\bar{a}_{*}$) and a short spike on a fuller reservoir ($\bar{a}>\bar{a}_{*}$). 
This fold is the spike analogue of the mass window \eqref{PCP_window} for the pinned front. 
Figure~\ref{AB_spike} compares the prediction \eqref{PCP_spike_profile} with a direct simulation of the triplet, which relaxes onto the tall branch; the measured height and width agree with \eqref{PCP_spike_profile} to within the finite-$D_{a}$ correction.
 
\paragraph{Separation of height, width, and spacing.} 
The spike profile \eqref{PCP_spike_profile} cleanly separates the three observables of a punctum.
Its \emph{height} $c_{\max}=3/(2\alpha\bar{a}\bar{b})$ is set by the feedback strength $\alpha$ and the conserved masses (through the reservoirs $\bar{a}$, $\bar{b}$), and is therefore tunable: increasing the available mass, or either monomer pool, raises the peak. 
Its \emph{width}, by contrast, is universal. 
The $\operatorname{sech}^{2}\!\big((x-x_{0})/2\big)$ shape carries no free constant, so the full width at half maximum (FWHM) is the parameter-free value $4\,\mathrm{arccosh}\sqrt{2}\approx 3.53$ in scaled units, \emph{independent} of $\alpha$, of the reservoirs, and of the masses -- the same quantities that move the height leave the width fixed. 
Restoring dimensions through $\mathcal{L}=\sqrt{D_{C}/v_{0}}$, the physical width is $\approx 3.53\,\sqrt{D_{C}/v_{0}}$, governed by the complex diffusivity $D_{C}$ and unbinding rate $v_{0}$ alone. 
Crucially the monomer diffusivities $D_{a}$, $D_{b}$ do not enter the single-spike width; they instead set the screening lengths $\sqrt{D_{a}}$, $\sqrt{D_{b}}$ over which a spike depletes its surroundings, and hence control the \emph{spacing} and number of puncta rather than their individual size (Sections~\ref{PCP_competition}--\ref{PCP_asym}). 
A punctum's amplitude is thus a readout of local mass and feedback, its footprint a readout of complex mobility, and the density of puncta along the junction a readout of monomer mobility.
\paragraph{Role of the heterodimer.} 
Two features distinguish the heterodimer (AB) spike from its homodimer (AA) counterpart of Section~\ref{PCP_homodimer}. 
First, the height \eqref{PCP_spike_profile} depends on the \emph{product} $\bar{a}\bar{b}$, so a punctum may be tuned through either monomer pool. 
Second, and more consequentially, the model carries two conserved reservoirs rather than one.
Subtracting the conservation laws, $a-b^{\dagger}$ obeys $\partial_{t}(a-b^{\dagger})=D_{a}a_{xx}-D_{b}b^{\dagger}_{xx}$, which is a pure diffusion equation \emph{only} when $D_{a}=D_{b}$; in that case $a-b^{\dagger}$ relaxes to the constant $a_{T}-b^{\dagger}_{T}$, the two reservoirs lock together, and the construction collapses to the single-reservoir (AA) problem. 
The genuinely two-reservoir behaviour of the heterodimer therefore emerges only for unequal monomer diffusivities $D_{a}\neq D_{b}$ (as in Figure~\ref{AB_onset_numerics}(b),(d)), where the two monomers possess distinct screening lengths $\sqrt{D_{a}}$ and $\sqrt{D_{b}}$.
 
\paragraph{From one punctum to many.} 
We have now characterised an individual punctum in full -- its existence threshold, its amplitude, and the clean separation of height, width and spacing -- and shown that on the tall branch a single spike is stable. 
A cell--cell junction, however, typically carries several puncta drawing on the same finite monomer pools, and there is no guarantee that such a configuration persists. 
In the next section we therefore ask whether a periodic train of spikes is stable, how it evolves if not, and what ultimately sets the number of puncta that survive.
 
\section{Arrays of puncta: number and robustness} \label{sec:arrays}

\subsection{Competition and coarsening} \label{PCP_competition}
 
A cell--cell junction typically carries more than one punctum distributed along its length~\citep{strutt2011dynamics, strutt2016robust}, so we now ask whether a periodic array of $K$ spikes is stable.

The calculation that follows has three steps, and it may help to state them before the algebra. 
First, an \emph{inner} problem: near each spike the complex perturbation responds to a monomer level that is essentially constant across the spike, which fixes the local response in terms of a single number $\eta_{j}$ per spike. 
Second, a \emph{global} coupling: each spike acts as a point source in the monomer field, and because the monomers are conserved that field is governed by a decay-free Green's function, so the $\eta_{j}$ are determined by all the spikes at once through a Green's matrix $\mathcal{G}$. 
Third, closing the loop between the two gives a scalar dispersion relation \eqref{PCP_disp_spike} in which the geometry of the array enters only through the eigenvalues $\mu_{m}$ of $\mathcal{G}$; the sign of $\mu_{m}$ then decides stability, and its magnitude sets the rate. 
The reader willing to take \eqref{PCP_disp_spike} on trust may skip to \eqref{PCP_lambda_comp}, which is the result used throughout the rest of the paper.

We linearise the decoupled unprimed triplet $(a,b^{\dagger},c)$ (the primed triplet $(a^{\dagger},b,c^{\dagger})$ behaves identically) about the $K$-spike steady state, writing $c=c_{E}+\mathrm{e}^{\lambda t}\psi$, $a=\bar{a}+\mathrm{e}^{\lambda t}\eta$, $b^{\dagger}=\bar{b}+\mathrm{e}^{\lambda t}\zeta$. In the symmetric case $\bar{a}=\bar{b}$, $D_{a}=D_{b}=D$ (so that $\eta=\zeta$) the linearisation of \eqref{a**:eq} and \eqref{c**:eq} reads
\begin{align}
\lambda\psi &= \psi_{xx} + (R_{c}-1)\psi + (R_{a}+R_{b^{\dagger}})\eta, \label{PCP_lin_c}\\
\lambda\eta &= D\eta_{xx} - (R_{c}-1)\psi - (R_{a}+R_{b^{\dagger}})\eta, \label{PCP_lin_a}
\end{align}
with $R_{c}=2\alpha c_{E}\bar{a}\bar{b}$, $R_{a}=\alpha c_{E}^{2}\bar{b}$ and $R_{b^{\dagger}}=\alpha c_{E}^{2}\bar{a}$ evaluated on the spike.
 
\paragraph{Local problem.} 
Near spike $j$, with $c_{E}=(\alpha\bar{a}^{2})^{-1}w$ from \eqref{PCP_spike_profile}, one finds $R_{c}=2w$ and $R_{a}+R_{b^{\dagger}}=2w^{2}/(\alpha\bar{a}^{3})$, while $\eta\approx\eta_{j}$ is constant across the $O(1)$-wide spike. 
The complex perturbation thus obeys
\begin{equation}
(\mathcal{L}_{0}-\lambda)\psi = -\beta\,\eta_{j}\,w^{2}, \qquad
\mathcal{L}_{0} := \partial_{yy}-1+2w, \quad \beta := \frac{2}{\alpha\bar{a}^{3}},
\label{PCP_local}
\end{equation}
where $\mathcal{L}_{0}$ is the standard spike operator. 
Differentiating the profile equation $w''-w+w^{2}=0$ yields the identities
\begin{equation}
\mathcal{L}_{0}w = w^{2}, \qquad
\mathcal{L}_{0}\!\left(w+\tfrac{1}{2}yw'\right)=w,
\label{PCP_identities}
\end{equation}
so that $\mathcal{L}_{0}^{-1}w^{2}=w$ and, using $\int w=\int w^{2}=6$,
\begin{equation}
P(\lambda) := \int_{-\infty}^{\infty}(\mathcal{L}_{0}-\lambda)^{-1}w^{2}\,\mathrm{d}y,
\qquad P(0)=6, \quad P'(0)=3 .
\label{PCP_Plambda}
\end{equation}
 
\paragraph{Global coupling.} 
Integrating \eqref{PCP_lin_c} across spike $j$ (the boundary terms vanish since $\psi$ is localised) gives the net reactive flux feeding the monomer field,
\begin{equation}
S_{j} := \int\!\big[(R_{c}-1)\psi + (R_{a}+R_{b^{\dagger}})\eta\big]\,\mathrm{d}y
= \lambda\!\int\!\psi\,\mathrm{d}y .
\label{PCP_source}
\end{equation}
The monomer perturbation then satisfies $D\eta_{xx}-\lambda\eta=\sum_{k}S_{k}\delta(x-x_{k})$ with
Neumann conditions. 
Crucially this operator carries \emph{no decay term} -- the monomers are conserved -- so its Green's function belongs to the near-shadow class \citep{mckay2012stability}, rather than the exponentially screened Green's function of activator--inhibitor models~\citep{iron2001stability}. 
Writing $\eta_{j}=\sum_{k}\mathcal{G}_{jk}S_{k}$ and using $\int\psi_{j}=-\beta\eta_{j}P(\lambda)$ from \eqref{PCP_local}, each eigenvector of the Green's matrix $\mathcal{G}$ (eigenvalue $\mu_{m}$) gives the dispersion relation
\begin{equation}
\lambda\,P(\lambda) = -\frac{\alpha\bar{a}^{3}}{2\mu_{m}} .
\label{PCP_disp_spike}
\end{equation}
 
\begin{figure}[htbp]
\centering
\includegraphics[width=\textwidth]{ 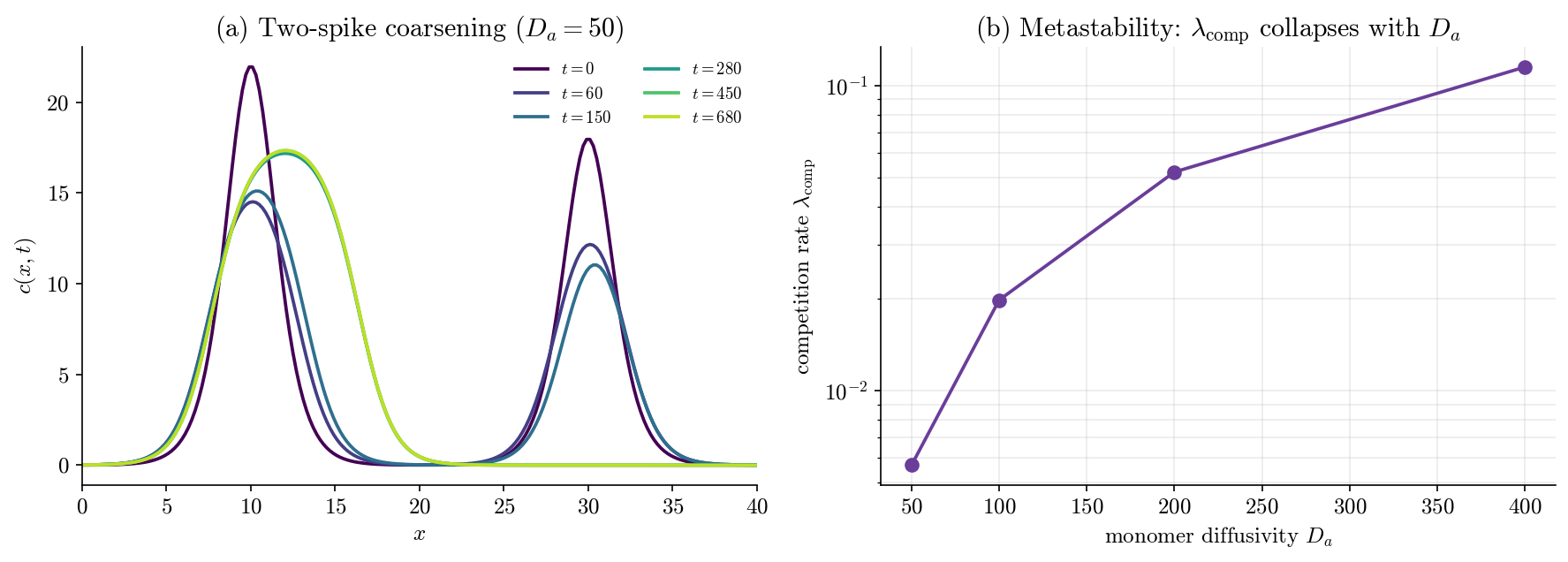}
\caption{\textbf{Competition of puncta.} 
(a) Two spikes seeded with a $10\%$ amplitude asymmetry ($\mathcal{K}(c)=c^{2}$, $\mathcal{V}=1$, $\alpha=0.3$, $D_{a}=D_{b}=50$, $\ell=40$): the larger spike grows at the expense of the smaller, which is extinguished by $t\approx280$, leaving a single punctum that then drifts slowly. 
(b) Competition rate $\lambda_{\mathrm{comp}}$, measured from the first $e$-folding of the spike-mass difference of a weakly perturbed pair, increasing with the monomer diffusivity as predicted by \eqref{PCP_lambda_comp} ($\lambda_{\mathrm{comp}}\propto D_{a}$, saturating at large $D_{a}$); the rate vanishes as $D_{a}\to0$, the metastable limit.
}
\label{AB_competition}
\end{figure}

\paragraph{Competition is generic.} 
The synchronous eigenvector $(1,\dots,1)$ carries the large Green's eigenvalue $\mu\sim-1/(\lambda\ell)$ forced by conservation; through \eqref{PCP_disp_spike} it corresponds to a stable amplitude mode (the same calculation selects the tall branch of \eqref{PCP_spike_sym} as the stable single spike). 
The dangerous modes are the \emph{competition} eigenvectors with $\sum_{k}(\cdot)_{k}=0$, whose eigenvalues $\mu_{m}$ are set by the regular part of the conserved Green's function. 
For these $\mu_{m}<0$ -- for two spikes $\mu=\mathcal{G}_{\mathrm{self}}-\mathcal{G}_{\mathrm{cross}}<0$, because the Neumann Green's function dips below its mean at the source -- and linearising \eqref{PCP_disp_spike} about $\lambda=0$ with $P(0)=6$ gives
\begin{equation}
\lambda_{\mathrm{comp}} \simeq -\frac{\alpha\bar{a}^{3}}{12\,\mu_{m}} \;>\;0 .
\label{PCP_lambda_comp}
\end{equation}
Every array with $K\ge 2$ is therefore linearly unstable: mass drains from smaller to larger spikes and the pattern coarsens toward a single punctum. 
Figure~\ref{AB_competition}(a) confirms this directly -- two spikes with a $10\%$ amplitude asymmetry collapse to one.

This conclusion is expected. 
Uninterrupted coarsening to a single peak is the generic behaviour of mass-conserving reaction--diffusion systems with one conserved species~\citep{otsuji2007mass, ishihara2007transient, brauns2021wavelength}, and the heterodimer inherits it: each triplet carries its own conserved pair, and within a triplet the argument above is the familiar one. 
What the calculation adds is not the sign of the instability but its \emph{coefficient} -- the closed form \eqref{PCP_rate_explicit}, and its two-reservoir generalisation \eqref{PCP_harmonic}, which is where the heterodimer ceases to behave like a single-reservoir system. 
The two-reservoir rate of Section~\ref{PCP_asym} is the substantive result.

\paragraph{Sign of the competition eigenvalue.} 
The instability claimed above rests on $\mu_{m}<0$ for every competition eigenvector, which we now establish in general rather than verify case by case. 
The relevant matrix is $H_{jk}=H(x_{j};x_{k})$, the regular part of the conserved Neumann Green's function \eqref{PCP_green} evaluated on the spike positions, and the competition subspace is the set of source vectors $S$ with $\sum_{k}S_{k}=0$ (the synchronous direction $(1,\dots,1)$ carrying instead the conserved amplitude mode). 
For such an $S$ define the field $\eta(x)=\sum_{k}S_{k}H(x;x_{k})$. 
By the defining equation $H''=\delta(x-x_{0})-1/\ell$ and $\sum_{k}S_{k}=0$, the uniform sink cancels and $\eta''=\sum_{k}S_{k}\delta(x-x_{k})$; multiplying by $\eta$, integrating over $(0,\ell)$ and using the Neumann conditions $\eta'(0)=\eta'(\ell)=0$ gives
\begin{equation}
S^{\top}H\,S \;=\; \sum_{j,k}S_{j}S_{k}\,H(x_{j};x_{k})
\;=\; \int_{0}^{\ell}\eta\,\eta''\,\mathrm{d}x
\;=\; -\int_{0}^{\ell}(\eta')^{2}\,\mathrm{d}x \;\le\;0,
\label{PCP_negdef}
\end{equation}
with equality only if $\eta'\equiv0$, i.e.\ $S=0$. Hence $H$ is negative definite on the competition subspace: every competition eigenvalue of $\mathcal{G}=H/D$ obeys $\mu_{m}<0$ for any number and arrangement of spikes, so by \eqref{PCP_lambda_comp} $\lambda_{\mathrm{comp}}>0$ for all admissible parameters. 
The two-spike value $\mu_{m}=-\ell/(4D)$ below is the simplest instance.
The argument is purely geometric -- it uses only the conservative (decay-free) structure of the monomer operator -- and so carries over unchanged to the asymmetric case: the eigenvalues of $\tfrac{1}{\bar{a}}\mathcal{G}_{a}+\tfrac{1}{\bar{b}}\mathcal{G}_{b}$ in \eqref{PCP_disp_asym} are $\tfrac{1}{\bar{a}}\tfrac{\nu}{D_{a}}+\tfrac{1}{\bar{b}}\tfrac{\nu}{D_{b}}$ with $\nu<0$ the (common) competition eigenvalue of $H$ from \eqref{PCP_negdef}, a positive combination of negative terms, so $\mu_{m}<0$ and the harmonic-mean rate \eqref{PCP_harmonic} is positive throughout. 
The only edge case is the spike-dissolution limit $D_{a}/D_{b}\approx10^{2}$ of Section~\ref{PCP_asym}: there the spike amplitude $c_{\max}=3/(2\alpha\bar{a}\bar{b})\to0$ and the reduction underlying \eqref{PCP_disp_asym} ceases to apply, but $\mu_{m}<0$ continues to hold right up to that limit -- competition does not change sign; rather, the punctum that would coarsen simply disappears.
 
\begin{figure}[htbp]
\centering
\includegraphics[width=\textwidth]{ 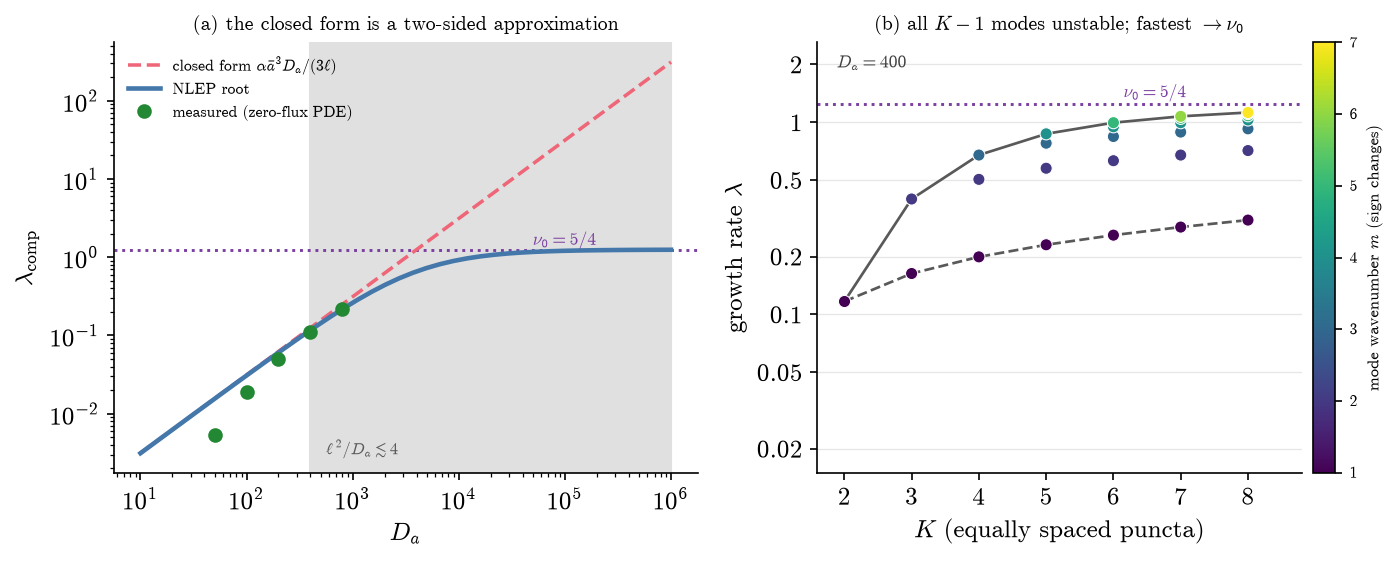}
\caption{\textbf{Validity of the closed-form competition rate, and its dependence
on punctum number.} 
The competition NLEP \eqref{PCP_disp_spike} rooted spectrally ($\mathcal{K}=c^{2}$, $\mathcal{V}=1$, $\alpha=0.3$, $\ell=40$, $a_{T}=b^{\dagger}_{T}=4.5$).
(a) Two spikes at $\ell/4,3\ell/4$. 
The closed form \eqref{PCP_rate_explicit} (dashed) is the $\lambda\to0$ linearisation and fails as $\lambda$ grows: the exact root (solid) leaves it beyond $D_{a}\sim10^{3}$ and saturates at $\nu_{0}=5/4$, while measured rates (circles) fall below both at small $D_{a}$, where the shadow limit fails. 
Agreement is therefore best over an intermediate window, which contains the band $\ell^{2}/D_{a}\lesssim4$ of Figures~\ref{AB_competition} and~\ref{AB_tworeservoir}.
(b) The $K-1$ competition eigenvalues of $H(x_{j};x_{k})$ for $K$ equally spaced puncta at $D_{a}=400$, coloured by number of sign changes $m$; grey curves trace the fastest ($m=K-1$, neighbour exchange) and slowest ($m=1$, end-to-end). 
All are unstable, as \eqref{PCP_negdef} requires, and the fastest approaches $\nu_{0}$ as $\ell/K$ shrinks --- a two-punctum array coarsens an order of magnitude more slowly than an eight-punctum one. 
Totals are fixed, so $\bar{a}$ rises with $K$ ($0.50\to1.08$).}
\label{AB_competition_saturation}
\end{figure}

\paragraph{Closed form for two spikes.} 
The regular part of the conserved Neumann Green's function is known explicitly~\citep{mckay2012stability}: $\mathcal{G}_{\mathrm{reg}}(x;x_{0})=H(x;x_{0})/D$ with
\begin{equation}
H(x;x_{0}) = \frac{|x-x_{0}|}{2} + \frac{x+x_{0}}{2} - \frac{x^{2}+x_{0}^{2}}{2\ell} - \frac{\ell}{3},
\label{PCP_green}
\end{equation}
the solution of $H''=\delta(x-x_{0})-1/\ell$ with $H'(0)=H'(\ell)=0$ and $\int_{0}^{\ell}H\,\mathrm{d}x=0$.
For two spikes at $\ell/4$ and $3\ell/4$ this gives $H_{\mathrm{self}}=-7\ell/48$ and $H_{\mathrm{cross}}=5\ell/48$, so the competition eigenvalue is $\mu_{m}=(H_{\mathrm{self}}-H_{\mathrm{cross}})/D=-\ell/(4D)$ and \eqref{PCP_lambda_comp} becomes the parameter-free rate
\begin{equation}
\lambda_{\mathrm{comp}} = \frac{\alpha\bar{a}^{3}D}{3\ell}.
\label{PCP_rate_explicit}
\end{equation}
This is the shadow-limit ($D_{a}\to\infty$) rate; the measured rate of Figure~\ref{AB_competition}(b) approaches it as $\ell^{2}/D_{a}\to0$ and agrees to within about ten percent in that regime (e.g.\ $0.115$ against the predicted $0.125$ at $D=400$, $\ell=40$, $\alpha=0.3$, $\bar{a}=0.5$).
Because \eqref{PCP_rate_explicit} is the $\lambda\to0$ linearisation of \eqref{PCP_disp_spike}, its accuracy is confined to an intermediate window in $D_{a}$: Figure~\ref{AB_competition_saturation}(a) compares it with the exact spectral root of the NLEP and with the measured rates, which fall below both at small $D_{a}$, where the shadow limit fails, while the closed form departs from the exact root beyond $D_{a}\sim10^{3}$, where $\lambda$ is no longer small.
For $K$ spikes the geometric factors are the eigenvalues of the explicit matrix $H(x_{j};x_{k})$.
All $K-1$ of these are unstable, as \eqref{PCP_negdef} requires, and the fastest -- exchange between neighbours -- accelerates as the puncta crowd together, so that an eight-punctum array coarsens an order of magnitude faster than a two-punctum one (Figure~\ref{AB_competition_saturation}(b)).
 
\paragraph{Metastability and the number of puncta.} The rate \eqref{PCP_lambda_comp} measures how strongly neighbouring spikes communicate through the shared monomer field. 
Since the conserved Green's function scales as $D^{-1}$, the competition eigenvalue obeys $|\mu_{m}|\propto D^{-1}$, so $\lambda_{\mathrm{comp}}\propto D_{a}$ at moderate diffusivity, saturating towards the local rate $\nu_{0}=\tfrac{5}{4}$ as $D_{a}\to\infty$. 
Figure~\ref{AB_competition}(b) confirms this: the measured rate rises monotonically with $D_{a}$. 
The mechanism is transparent -- coarsening proceeds by transporting monomer between spikes, and faster monomer diffusion accelerates that transport. 
Conversely $\lambda_{\mathrm{comp}}\to0$ as $D_{a}\to0$, so multi-punctum states become \emph{metastable}: weakly diffusing monomers leave neighbouring puncta effectively decoupled, and an array persists for a long time. 
There is thus no sharp linear-stability threshold of the kind found for activator--inhibitor spikes \citep{iron2001stability} -- every array with $K\ge2$ is formally unstable -- but the slowness of coarsening at small $D_{a}$ plays the same role in practice. 

The number of puncta surviving over the polarisation timescale is therefore set kinetically: a spike depletes its surroundings over the monomer screening length $\sqrt{D_{a}}$, so spikes closer than $\sim\!\sqrt{D_{a}}$ compete strongly and a long-lived array carries at most $\sim\!\ell/\sqrt{D_{a}}$ puncta. 
The same logic -- formal instability rendered irrelevant by the slowness of competition, so that the observed number reflects kinetics rather than a selected wavelength -- was established for Rho-GTPase polarity sites by \citet{chiou2018principles, chiou2021how}, there with the conserved load rather than the monomer mobility as the control parameter; the mechanism is the same, and we claim no priority for it. 
Smaller monomer diffusivity thus favours more, and more persistent, puncta -- both because the screening length is shorter and because coarsening is slower. 
The asymmetric-diffusion case $D_{a}\neq D_{b}$, and the oscillatory modes admitted by the three-field structure, are taken up in Section~\ref{PCP_asym}.

\begin{figure}[htbp]
\centering
\includegraphics[width=\textwidth]{ 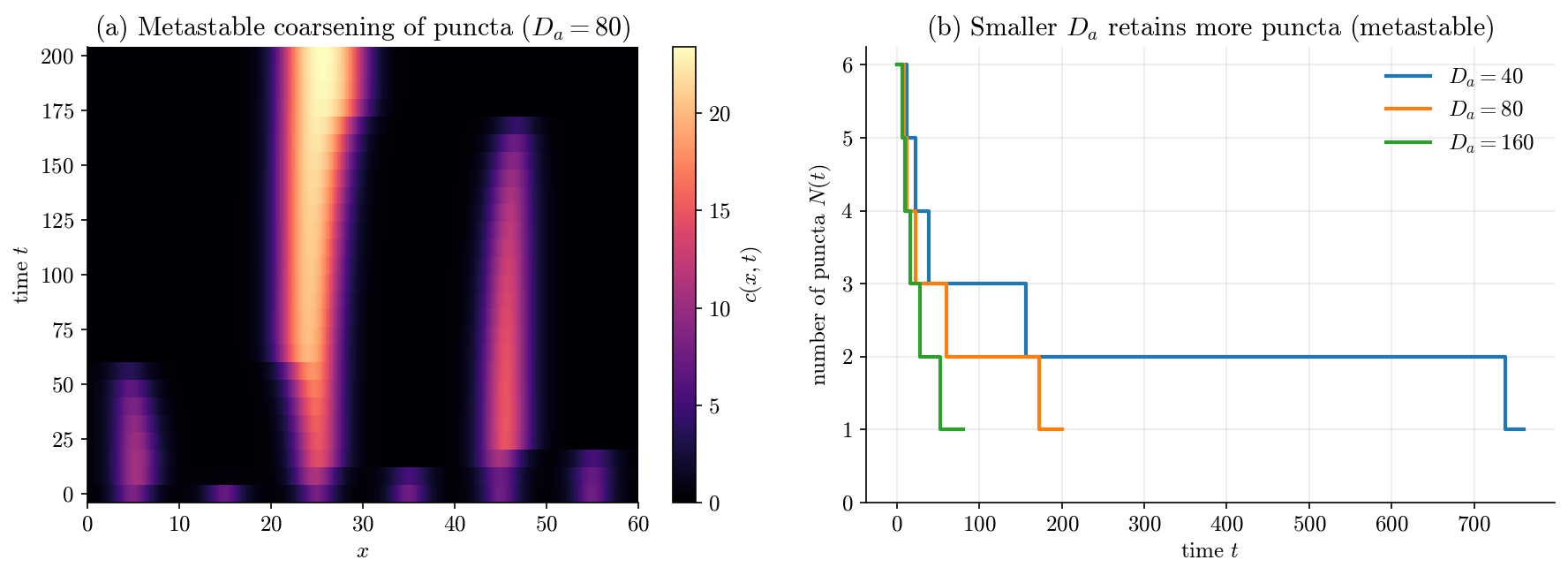}
\caption{\textbf{Metastable coarsening of a punctum array} ($\mathcal{K}(c)=c^{2}$, $\mathcal{V}=1$, $\alpha=0.3$, $\ell=60$, $a_{T}=b^{\dagger}_{T}=3.9$). 
(a) Space--time plot of $c(x,t)$ for $D_{a}=D_{b}=80$ seeded with six spikes: the array coarsens stepwise to a single drifting punctum, with progressively longer plateaus. 
(b) Number of puncta $N(t)$ from the same initial condition for $D_{a}=40,80,160$; smaller $D_{a}$ coarsens more slowly and retains more puncta -- the kinetic counterpart of \eqref{PCP_lambda_comp}.}
\label{AB_coarsening}
\end{figure}

\paragraph{Coarsening of a multi-punctum array.} The pairwise picture extends to a full array.
Figure~\ref{AB_coarsening}(a) follows six spikes seeded on a cell--cell junction: the pattern coarsens in discrete steps, each surviving spike absorbing a neighbour and drifting slowly, with the waiting time between events lengthening as the puncta grow farther apart -- the hallmark of metastable coarsening.
Figure~\ref{AB_coarsening}(b) tracks the number of puncta $N(t)$ for three monomer diffusivities from the same initial configuration: consistent with $\lambda_{\mathrm{comp}}\propto D_{a}$, the array coarsens fastest at large $D_{a}$ ($N$ reaches unity by $t\approx80$ for $D_{a}=160$) and slowest at small $D_{a}$ (two puncta persist to $t\approx750$ for $D_{a}=40$). 
Over a fixed polarisation window, a cell-cell junction with less mobile monomers therefore supports more long-lived puncta.
 
\subsection{Asymmetric diffusion and oscillatory modes} \label{PCP_asym}
 
\begin{figure}[htbp]
\centering
\includegraphics[width=\textwidth]{ 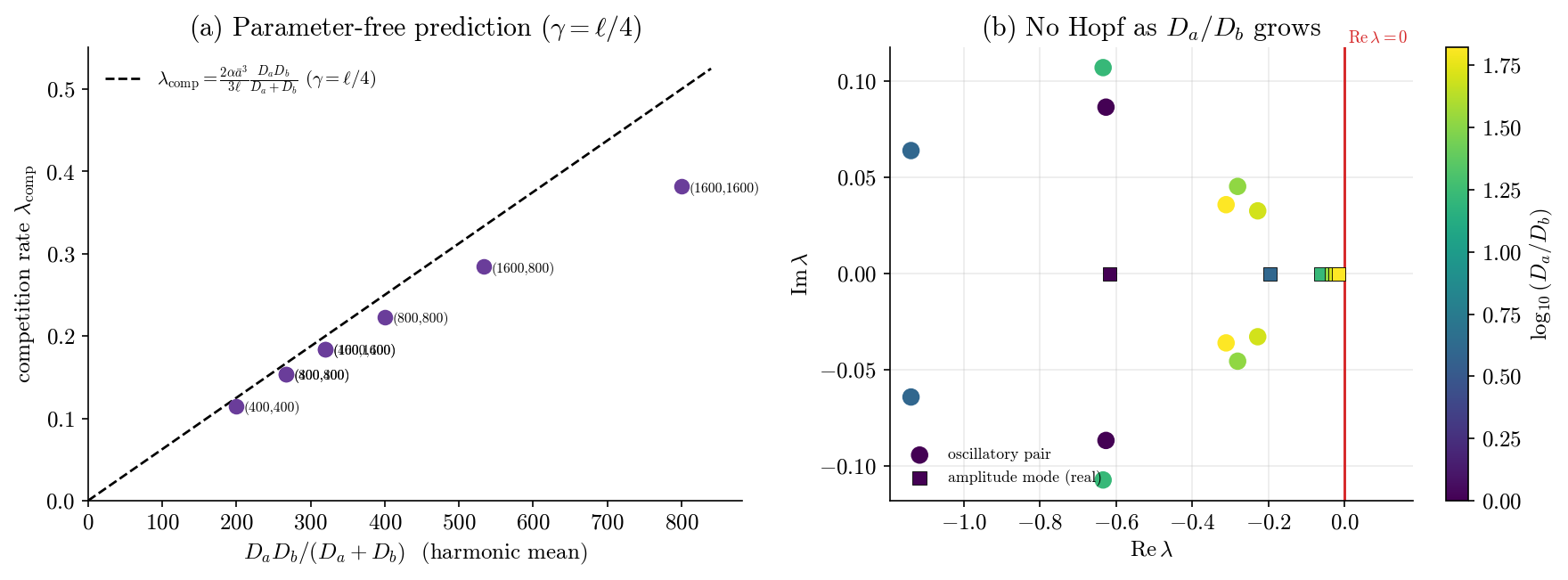}
\caption{\textbf{Two-reservoir effects.} 
(a) Competition rate $\lambda_{\mathrm{comp}}$, measured from the spike-mass $e$-folding of a perturbed pair, against the harmonic mean $D_{a}D_{b}/(D_{a}+D_{b})$ for eight $(D_{a},D_{b})$ combinations with $D_{a},D_{b}\ge400$ ($a_{T}=b^{\dagger}_{T}$, so $\bar{a}=\bar{b}$); the collapse onto a straight line through the origin confirms the harmonic-mean scaling \eqref{PCP_harmonic} (the points lying just below the shadow-limit line by the finite-$D_{a}$ correction), and $(D_{a},D_{b})\leftrightarrow (D_{b},D_{a})$ pairs coincide.
(b) Single-spike spectrum swept over the diffusivity ratio $D_{a}/D_{b}$ (colour; $D_{a}=100$): the amplitude mode (squares) stays real and tends to $0^{-}$ as the spike dissolves near $D_{a}/D_{b}\approx10^{2}$, while the oscillatory pairs (circles) remain at $\mathrm{Re}\,\lambda\le-0.2$; none crosses the axis (red), so the punctum is a damped focus, not a Hopf oscillator.}
\label{AB_tworeservoir}
\end{figure}

\begin{figure}[htbp]
\centering
\includegraphics[width=\textwidth]{ 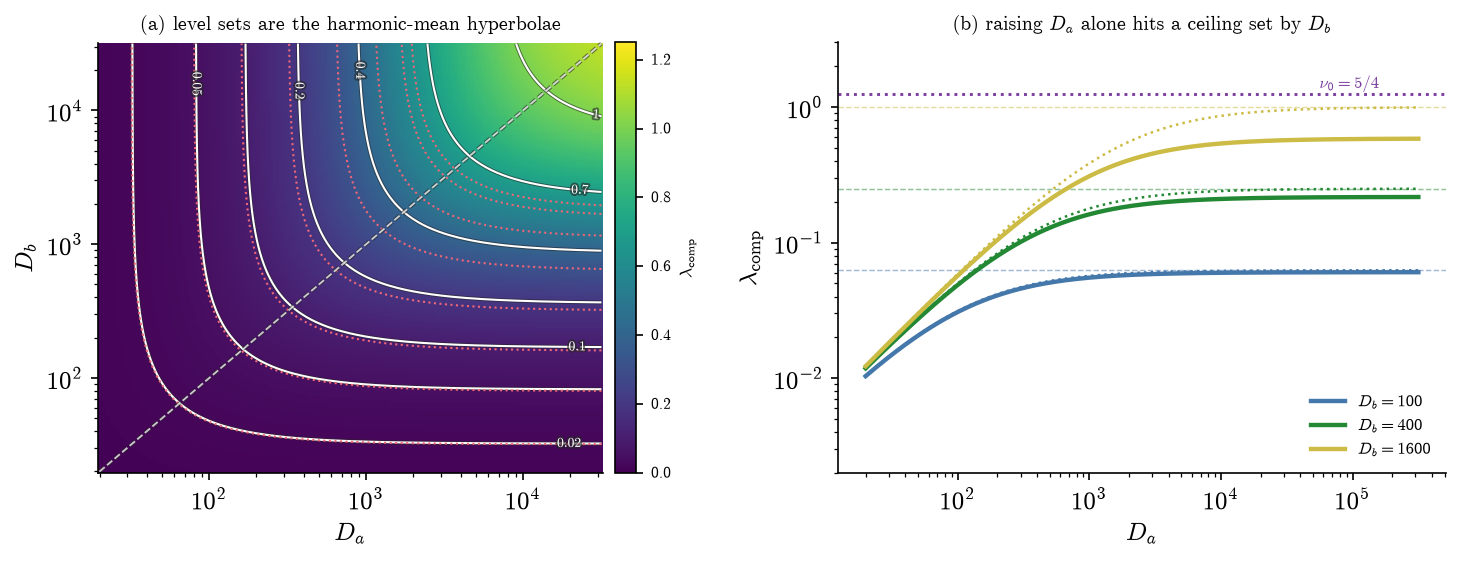}
\caption{
\textbf{The two reservoirs enter through their harmonic mean, so the slower monomer sets the rate.} 
The asymmetric competition NLEP \eqref{PCP_disp_asym} rooted spectrally for two spikes at $\ell/4,3\ell/4$ ($\gamma=\ell/4$, $\mathcal{K}=c^{2}$, $\mathcal{V}=1$, $\alpha=0.3$, $\ell=40$, $a_{T}=b^{\dagger}_{T}=4.5$, $\bar{a}=\bar{b}=0.5$). (a) $\lambda_{\mathrm{comp}}$ over the $(D_{a},D_{b})$ plane (colour, white contours), with the closed form \eqref{PCP_harmonic} at the same levels (blue dotted) and the diagonal $D_{a}=D_{b}$ (grey dashed). 
The level sets are the hyperbolae $1/(\bar{a}D_{a})+1/(\bar{b}D_{b})=\mathrm{const}$ and are symmetric about the diagonal, as \eqref{PCP_harmonic} requires when $\bar{a}=\bar{b}$; each contour turns parallel to the axes, so once one monomer is fast, making it faster buys nothing. 
The closed form tracks the exact contours at small diffusivity and over-predicts at large, where $\lambda$ is no longer small.
(b) $\lambda_{\mathrm{comp}}$ against $D_{a}$ at fixed $D_{b}$ (exact solid, closed form dotted). 
Each curve flattens at a ceiling fixed by $D_{b}$ alone, $2\alpha\bar{a}^{3}D_{b}/(3\ell)$ in the linearised law (thin dashed) --- twice the equal-diffusivity rate at $D=D_{b}$. 
The exact ceilings are $0.061$, $0.217$ and $0.586$ for $D_{b}=100$, $400$, $1600$, against $0.063$, $0.250$ and $1.000$ from \eqref{PCP_harmonic}: the closed form is accurate while the ceiling is well below $\nu_{0}=5/4$ and fails as it approaches it.
}
\label{AB_tworeservoir_plane}
\end{figure}

\paragraph{Two reservoirs.} 
When $D_{a}\neq D_{b}$ the monomer perturbations $\eta$ and $\zeta$ no longer coincide, but they are sourced by the \emph{same} localised flux $S_{j}=\lambda\int\psi_{j}$, because each spike consumes both monomers together. 
Hence $\eta=\mathcal{G}_{a}S$ and $\zeta=\mathcal{G}_{b}S$ with the two substrate Green's matrices, and the dispersion relation \eqref{PCP_disp_spike} generalises to
\begin{equation}
\lambda\,P(\lambda) = -\frac{\alpha\bar{a}\bar{b}}{\mu_{m}}, \qquad \mu_{m} = \mathrm{eig}\!\left(\tfrac{1}{\bar{a}}\mathcal{G}_{a}+\tfrac{1}{\bar{b}}\mathcal{G}_{b}\right).
\label{PCP_disp_asym}
\end{equation}
Since $\mathcal{G}_{a}$ and $\mathcal{G}_{b}$ are built on the same spike positions they share eigenvectors, with competition eigenvalues $-\gamma/D_{a}$ and $-\gamma/D_{b}$; the geometric factor is fixed in closed form by \eqref{PCP_green}, namely $\gamma=\ell/4$ for two spikes at $\ell/4,3\ell/4$. 
The competition rate is therefore
\begin{equation}
\lambda_{\mathrm{comp}} = \frac{\alpha(\bar{a}\bar{b})^{2}}{6\gamma}\,
\frac{D_{a}D_{b}}{\bar{a}D_{a}+\bar{b}D_{b}}
\;\xrightarrow[\ \gamma=\ell/4\ ]{\ \bar{a}=\bar{b}\ }\;
\frac{2\alpha\bar{a}^{3}}{3\ell}\,\frac{D_{a}D_{b}}{D_{a}+D_{b}} .
\label{PCP_harmonic}
\end{equation}
The rate scales with the \emph{harmonic mean} of the two diffusivities and is thus rate-limited by the \emph{less} mobile monomer: if either reservoir cannot be transported between spikes, mass cannot redistribute.
Figure~\ref{AB_tworeservoir}(a) confirms the law \emph{with no free parameter}: the eight $(D_{a},D_{b})$ rates (here with $D_{a},D_{b}\ge400$, so that $\ell^{2}/D_{a}\lesssim4$) fall just below the parameter-free shadow-limit line $\lambda_{\mathrm{comp}}=(2\alpha\bar{a}^{3}/3\ell)\, D_{a}D_{b}/(D_{a}+D_{b})$, and $(D_{a},D_{b})\leftrightarrow(D_{b},D_{a})$ pairs coincide, as \eqref{PCP_harmonic} requires when $\bar{a}=\bar{b}$. 
Here the heterodimer departs qualitatively from a single-monomer model: a punctum is only as labile as its slowest partner. 
That an array coarsens, and that coarsening slows as transport slows, is inherited from the single-reservoir theory~\citep{otsuji2007mass, brauns2021wavelength}. 
What has no single-reservoir analogue is the \emph{form} of the dependence: because a spike consumes both monomers together but each is transported by its own Green's function, the two mobilities enter \eqref{PCP_harmonic} through their harmonic mean rather than additively, so the level sets of $\lambda_{\mathrm{comp}}$ are hyperbolae and each contour turns parallel to the axes (Figure~\ref{AB_tworeservoir_plane}(a)). 
A single-species model has no way to express this: it predicts a rate proportional to the one diffusivity available, and cannot produce a ceiling set by the other reservoir.
That ceiling is explicit in Figure~\ref{AB_tworeservoir_plane}(b): raising $D_{a}$ at fixed $D_{b}$ saturates at a value fixed by $D_{b}$ alone, twice the equal-diffusivity rate at $D=D_{b}$.
 
\paragraph{Oscillatory modes.} 
A three-field system can support oscillatory instabilities that a two-field one cannot \citep{mckay2012stability}, and the spectrum of the linearised single-spike operator indeed contains complex-conjugate pairs (Figure~\ref{AB_tworeservoir}(b)): a punctum is a focus, relaxing through a damped breathing motion. 
Two results show that these pairs never destabilise. 
First, in the shadow limit $D_{a},D_{b}\to\infty$ the Green's matrices collapse to their conserved zero mode, $\mathcal{G}\to-(\lambda\ell)^{-1}$, and \eqref{PCP_disp_asym} reduces to the real equation $P(\lambda)=\alpha(\bar{a}\bar{b})^{2}\ell/(\bar{a}+\bar{b})$, so the eigenvalue nearest the imaginary axis -- the amplitude mode -- is real. 
Second, sweeping the diffusivity ratio from $D_{a}/D_{b}=1$ up to the spike-existence limit $D_{a}/D_{b}\approx10^{2}$ (beyond which the spike dissolves, $c_{\max}\to0$), the only eigenvalue that approaches the axis is this real amplitude mode, which tends to $0^{-}$ as the spike disappears, while the oscillatory pairs stay bounded away at $\mathrm{Re}\,\lambda\le-0.2$ (Figure~\ref{AB_tworeservoir}(b)). 
The puncta therefore do not undergo a Hopf bifurcation. 
A kinetic timescale separation does not create one either: slowing the monomers ($\tau\,a_{t}=D_{a}a_{xx}-\dots$) drives a \emph{real} eigenvalue unstable -- a monotonic amplitude instability, not an oscillation -- while slowing the complex merely lengthens the relaxation.
Sustained oscillatory (``blinking'') puncta therefore lie outside the conserved heterodimer kinetics, consistent with the observed persistence and reduced turnover of core-protein puncta; a genuine Hopf would require non-conservative turnover or an explicit feedback delay.

\section{Sorting: segregation of oppositely oriented puncta for \texorpdfstring{$\mathcal{V}'>0$}{V'>0}} \label{PCP_polarity}
 
Every result so far holds at $\mathcal{V}'=0$, where \eqref{a**:eq}--\eqref{cdagger**:eq} splits into two independent triplets: puncta of both orientations form, compete and coarsen, but the model is blind to their \emph{relative} arrangement. 
The linear analysis of Section~\ref{PCP_dispersion} already shows that this degeneracy is broken by $\mathcal{V}'\neq0$, the antisymmetric (sorting) mode $\tilde{c}-\tilde{c}^{\dagger}$ losing stability first precisely when $\mathcal{V}'(\bar{c})>0$.
We now carry that statement to the fully nonlinear puncta.
 
The two triplets are coupled only through the unbinding rate: the $c$-equation carries $-\mathcal{V}(c^{\dagger})c$. 
Writing $\mathcal{V}(c^{\dagger})=\mathcal{V}(0)+\mathcal{V}'c^{\dagger}+\dots$ and absorbing $\mathcal{V}(0)$ into the unbinding timescale, the coupling contributes a term $-\mathcal{V}'c^{\dagger}c$ to the $c$-dynamics. 
A lone punctum sits where the opposite complex is absent, $c^{\dagger}\approx0$, and is therefore unperturbed; the coupling acts only where the two profiles overlap.

\paragraph{Drift law.} 
Treat $\mathcal{V}'$ as a weak coupling, so that to leading order $c=S(x-x_{c})$ is the spike \eqref{PCP_spike_profile}, $S=(\alpha\bar{a}\bar{b})^{-1}w$. 
The extra term $-\mathcal{V}'c^{\dagger}c$ drives a slow translation of the punctum. 
Projecting the perturbed equation onto the translation mode $S'$ (the kernel of the linearised spike operator), using $SS'=\tfrac{1}{2}(S^{2})'$ and integrating by parts gives
\begin{equation}
\dot{x}_{c} = -\frac{\mathcal{V}'}{2}
  \frac{\int c^{\dagger\prime}(x) S^{2}(x-x_{c}) \mathrm{d}x}
       {\int (S')^{2} \mathrm{d}x} .
\label{PCP_drift}
\end{equation}
What moves the punctum is therefore not the gradient of the opposite complex at its centre, but that gradient averaged across the punctum with weight $S^{2}$.
The two differ, because $c^{\dagger}$ is itself a $\operatorname{sech}^{2}$ punctum of the same universal width \eqref{PCP_spike_profile} and its gradient varies over the punctum on the punctum's own length scale. 
Collecting the mismatch into a single coefficient, $\dot{x}_{c}=-\chi(s)\mathcal{V}' c^{\dagger\prime}(x_{c})$ at a separation $s=x_{c^{\dagger}}-x_{c}$, with
\begin{equation}
\chi(s) = \frac{5}{2}
  \frac{\int c^{\dagger\prime}S^{2}}{c^{\dagger\prime}(x_{c})\int S^{2}}
  = \frac{5}{18}\frac{\Phi(s)}{\operatorname{sech}^{2}(s/2)\tanh(s/2)} ,
\label{PCP_chi_of_s}
\end{equation}
where $\Phi(s)=\int w'(y-s)w^{2}(y)\mathrm{d}y$ and $5/2=\int w^{2}/2\int(w')^{2}$ is the value $\chi$ would take if $c^{\dagger\prime}$ were constant over the punctum (for $w=\tfrac{3}{2}\operatorname{sech}^{2}(y/2)$, $\int w^{2}=6$, $\int(w')^{2}=6/5$ and $\int w(w')^{2}=36/35$), and symmetrically $\dot{x}_{c^{\dagger}}=-\chi(s)\mathcal{V}'c'(x_{c^{\dagger}})$.

Both ends of \eqref{PCP_chi_of_s} follow from that reading. 
Near contact the weight $S^{2}$ straddles the peak of $c^{\dagger}$, where $c^{\dagger\prime}$ changes sign, so the two halves oppose and the average falls below the central value: $\Phi(0)=0$, $\Phi'(0)=2\int w(w')^{2}$, and $\chi(0)=\tfrac{4}{3}\int w(w')^{2}/\int(w')^{2}=8/7$. 
Well separated, the punctum sits in the exponential tail $w\to6\mathrm{e}^{-|y|}$, where $c^{\dagger\prime}$ grows across it like $\mathrm{e}^{x}$, so the average exceeds the central value by $\int\mathrm{e}^{y}\operatorname{sech}^{4}(y/2)\mathrm{d}y/ \int\operatorname{sech}^{4}(y/2)\mathrm{d}y=(16/3)/(8/3)=2$ and $\chi(\infty)=5$.

\begin{figure}[htbp]
\centering
\includegraphics[width=\textwidth]{ 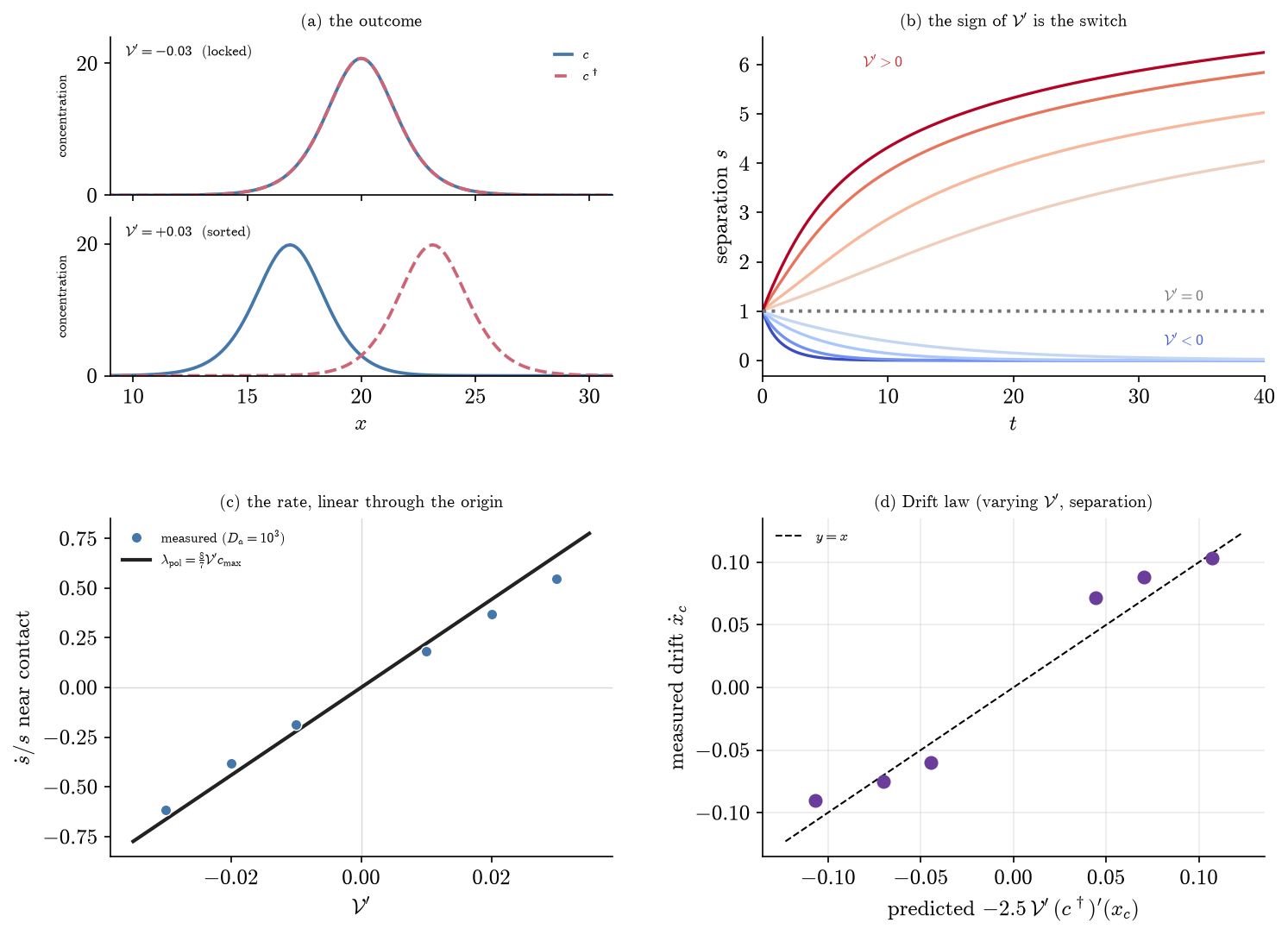}
\caption{\textbf{Sorting: the sign of $\mathcal{V}'$ decides whether co-located puncta segregate.} 
Two puncta, one in each triplet, coupled only through $\mathcal{V}(c^{\dagger})=1+\mathcal{V}'c^{\dagger}$, with fast monomers ($\alpha=0.3$, $\ell=40$, $n_{T}=2.5$, $c_{\max}\approx19.4$).
(a) Profiles at $t=40$ from the same initial separation at $D_{a}=10^{3}$: the two orientations are co-located for $\mathcal{V}'<0$ and disjoint for $\mathcal{V}'>0$.
(b) Separation against time for $\mathcal{V}'$ from $-0.03$ to $0.03$: trajectories diverge for $\mathcal{V}'>0$, collapse to contact for $\mathcal{V}'<0$, and are stationary at $\mathcal{V}'=0$ (dotted). Since $\chi(s)>0$, $s=0$ is the only equilibrium of the pair and the sign of $\mathcal{V}'$ alone fixes its stability.
(c) The near-contact rate $\dot{s}/s$ against $\mathcal{V}'$: linear through the origin for both signs, with the slope $\tfrac{8}{7}c_{\max}$ of \eqref{PCP_lambda_pol} and no free parameter.
(d) Measured punctum drift $\dot{x}_{c}$ against the prediction $-\tfrac{5}{2}\mathcal{V}'c^{\dagger\prime}(x_{c})$ of \eqref{PCP_drift}, over a range of separations and both signs of $\mathcal{V}'$ at $D_{a}=D_{b}=10^{3}$ (fitted slope $1.03$).
}
\label{AB_sorting}
\end{figure}

The coefficient therefore rises monotonically from $8/7$ at contact to $5$ when the puncta are well separated, and no constant reproduces both ends: $5/2$ exceeds $\chi$ by $35/16$ at contact and falls short of it by a factor of two in the far field, matching \eqref{PCP_chi_of_s} at the single separation $s=2.77$.
Since $\chi(s)>0$ throughout, each punctum drifts along the gradient of the \emph{opposite} complex: down-gradient (mutual repulsion) for $\mathcal{V}'>0$,
up-gradient (attraction) for $\mathcal{V}'<0$ (Figure~\ref{AB_sorting}(a)).
Because $\chi$ never vanishes, $s=0$ is the only equilibrium of the pair, and the sign of $\mathcal{V}'$ alone fixes its stability: separations diverge for $\mathcal{V}'>0$, collapse to contact for $\mathcal{V}'<0$, and are stationary at $\mathcal{V}'=0$ (Figure~\ref{AB_sorting}(b)).
Figure~\ref{AB_sorting}(d) tests the drift law \eqref{PCP_drift} directly, comparing the measured translation of a punctum with the prediction $-\tfrac{5}{2}\mathcal{V}'c^{\dagger\prime}(x_{c})$ across a range of separations and both signs of $\mathcal{V}'$; the fitted slope is $1.03$.
 
\paragraph{Spike-level sorting instability.} 
A co-located pair $x_{c}=x_{c^{\dagger}}$ is a symmetric fixed point of \eqref{PCP_drift}. 
Displacing the pair by $\pm\xi$ and expanding $c^{\dagger\prime}$ to first order, the separation $s=x_{c}-x_{c^{\dagger}}$ obeys $\dot{s}=\lambda_{\mathrm{pol}}s$ with
\begin{equation}
\lambda_{\mathrm{pol}} = \frac{2\mathcal{V}'}{\alpha\bar{a}\bar{b}}\,
\frac{\int w(w')^{2}}{\int(w')^{2}}
= \frac{12\,\mathcal{V}'}{7\,\alpha\bar{a}\bar{b}} = \frac{8}{7}\,\mathcal{V}'\,c_{\max},
\label{PCP_lambda_pol}
\end{equation}
using $\int w(w')^{2}=36/35$ and $c_{\max}=3/(2\alpha\bar{a}\bar{b})$. 
The co-located state is unstable -- the puncta segregate -- if and only if $\mathcal{V}'>0$, exactly the criterion of the linear antisymmetric mode, now realised by fully formed spikes; the rate is set by the punctum height, so taller puncta sort faster. 
Equation \eqref{PCP_lambda_pol} is the contact limit of the drift law \eqref{PCP_chi_of_s} rather than an independent result, so the two are consistent by construction. 
Numerically the instability has the predicted sign, is linear in $\mathcal{V}'$ through the origin for both signs, and grows in proportion to $c_{\max}$ (Figure~\ref{AB_sorting}(c)). 
The measured rate falls short of \eqref{PCP_lambda_pol} by $10$--$17\%$ at $D_{a}=10^{3}$ but by under $5\%$ at $D_{a}=10^{5}$; the deficit is the monomer's own odd response to the displaced punctum, a term the rigid-spike derivation drops, and it scales as $\ell^{2}/D_{a}$ rather than being a fixed property of the model.

\paragraph{Interpretation.} For $\mathcal{V}'>0$ -- each complex promoting the unbinding of the
complex of opposite orientation, the competitive cross-interaction attributed to Prickle in
Section~\ref{sec:introduction} -- a co-located pair cannot persist: the two species drift apart until
their overlap, and with it the coupling, is exhausted, leaving $c$ and $c^{\dagger}$ puncta in disjoint
regions. This is sorting generated by the punctum dynamics rather than inherited from the
initial data (contrast Figure~\ref{AB_onset_numerics}, where $\mathcal{V}'=0$ leaves placement to the
seeding noise). The opposite sign $\mathcal{V}'<0$ co-stabilises the two orientations and locks them
together (Figure~\ref{AB_sorting}(a)). The mass-redistribution machinery thus selects
\emph{how many} puncta form (Section~\ref{PCP_competition}); the sign of $\mathcal{V}'$ selects
\emph{whether they sort}.
 
\paragraph{Caveat.} $\mathcal{V}$ is a scalar caricature of the phosphorylation- and
endocytosis-mediated antagonism among the core components, and \eqref{PCP_drift}--\eqref{PCP_lambda_pol}
are weak-coupling ($\mathcal{V}'$ small) asymptotics. Within those limits they isolate the minimal
ingredient that turns clustering into sorting: a sign-definite cross-modulation of complex turnover,
with $\mathcal{V}'>0$ the sorting sign.

\paragraph{Scope of the sorting result.} The domain here is a single cell--cell contact,
so \eqref{PCP_drift}--\eqref{PCP_lambda_pol} say that two oppositely oriented puncta cannot occupy the
same region of that contact: the cross-modulation supplies \emph{mutual exclusion at the molecular
scale}. Cell-scale planar polarity is a different and stronger statement -- that a given contact comes
to be dominated by one orientation, and that opposite contacts of the same cell are dominated by
opposite orientations. That requires the contacts of a cell to be coupled through the reservoirs they
share, and a bias to select which orientation prevails where; neither is present in the single-contact,
zero-flux setting adopted here. What the present analysis establishes is therefore that
$\mathcal{V}'>0$ is sufficient for the local exclusion step that cell-scale models of the pathway
posit phenomenologically~\citep{amonlirdviman2005mathematical, burak2009order, fisher2019modelling},
and that this step follows from punctum dynamics rather than having to be imposed. The corresponding
prediction is sub-junctional: within a single contact, puncta of the two orientations should be
anticorrelated in position, with an exclusion zone set by the punctum width \eqref{PCP_spike_profile}
rather than by any imposed length scale -- a statement accessible to two-colour imaging of the two
orientations along a contact.
\paragraph{Robustness to the feedback form.} 
The drift law \eqref{PCP_drift} and the rate $\lambda_{\mathrm{pol}}$ \eqref{PCP_lambda_pol} are derived for the unbounded ($\mathcal{K}=c^{2}$) spike, but the segregation they describe is not an artefact of the spike regime. 
The criterion that selects sorting is already established at linear order in Section~\ref{PCP_dispersion}, where the antisymmetric (sorting) mode is the first to lose stability precisely when $\omega=\bar{c}\, \mathcal{V}'(\bar{c})>0$ \eqref{PCP_parity}. 
That calculation is carried out at the symmetric steady state for a \emph{general} clustering feedback $\mathcal{K}$: the parity splitting $\phi_{\pm}=\phi\pm\omega$ depends on the sign of $\mathcal{V}'(\bar{c})$ alone and not on whether $\mathcal{K}$ saturates. 
The saturating (mesa) and unbounded (spike) feedbacks therefore share the same sorting criterion $\mathcal{V}'>0$; what changes between them is only the morphology of the clustered state -- a pinned plateau rather than a tall peak -- and hence the quantitative prefactors in \eqref{PCP_drift}--\eqref{PCP_lambda_pol}, which are specific to the $\mathrm{sech}^{2}$ profile.
Sorting thus rests on the cross-modulation of turnover and is inherited by both morphologies; a direct simulation of a $\mathcal{V}'>0$ pair under the saturating feedback would provide a useful confirmation of this shared criterion.

\section{Discussion} \label{sec:discussion}
 
We set out to determine whether a cell junction carrying reversibly binding planar polarity complexes must always relax to a spatially uniform state, or whether it can spontaneously organise into puncta. 
Our analysis shows that a mass-conserving heterodimer model with concentration-dependent clustering feedback does pattern: above a threshold total density the uniform state loses stability to a long-wavelength, mass-redistribution mode (Section~\ref{sec:onset}), and the resulting localised structures are the puncta. 
The same instability underlies two distinct morphologies -- narrow spikes when the clustering feedback is unbounded, broad mesas when it saturates (Section~\ref{sec:single}) -- which differ in degree rather than in mechanism. 
Most existing planar polarity models resolve the cell or the tissue and treat each junction as well mixed~\citep{legarrec2006establishment, burak2009order, hazelwood2013functional, amonlirdviman2005mathematical, fischer2013persistent}; where subcellular structure is represented, it is through discrete compartments or lattice sites rather than as a continuum along the contact~\citep{thayambath2026start}. 
In contrast, the present framework resolves the puncta themselves, allowing us to ask what controls their size, number, and arrangement.

We now return to the questions posed in the Introduction. 
\emph{What minimal interactions establish a stable punctum?} 
Mass-conserving \emph{trans}-binding with concentration-dependent clustering feedback, above a threshold density, is sufficient (Sections~\ref{sec:onset}--\ref{sec:single}). 
\emph{One punctum or several?} 
A single punctum per conserved reservoir is the only true attractor, but coarsening is slow, so several coexist metastably; the number is selected kinetically rather than by a fixed wavelength (Section~\ref{PCP_competition}). 
\emph{How do size, shape, spacing and stability depend on the parameters?} 
The spike analysis separates these cleanly -- amplitude set by mass and feedback, a universal width set by the complex diffusion length, and spacing set by the monomer screening length (Section~\ref{PCP_spike}). 
\emph{Does the clustering machinery also drive sorting?} 
No: clustering and sorting are controlled by separate ingredients, sorting requiring in addition a sign-definite cross-modulation of turnover, $\mathcal{V}'>0$ (Sections~\ref{PCP_dispersion},~\ref{PCP_polarity}). 
We expand on each of these points below.
 
\paragraph{What sets the size and number of puncta.} 
The spike analysis cleanly separates the three observables of a punctum (Section~\ref{PCP_spike}). 
The amplitude is set by the local available mass and the feedback strength $\alpha$, and is therefore tunable through either monomer pool. 
The width is universal, fixed by the complex diffusion length alone and independent of the masses and of the feedback. 
The spacing, and hence the number of puncta, is governed by the monomer screening length $\sqrt{D_{a}}$. 
Because every array of two or more spikes is formally unstable and coarsens (Section~\ref{PCP_competition}), the number of puncta is selected not by a sharp linear-stability threshold but kinetically: coarsening slows as the monomers become less mobile, so the count surviving over the polarisation window reflects how far coarsening has progressed rather than a fixed wavelength. 
A concrete, testable prediction follows -- junctions with less mobile monomers should support more, and longer-lived, puncta.
 
\paragraph{Clustering versus sorting.} 
Clustering alone does not produce sorting. 
At $\mathcal{V}'=0$ the two complex orientations form and coarsen independently, and their relative placement is inherited from the initial data rather than selected (Figure~\ref{AB_onset_numerics}). 
Sorting emerges only when the unbinding of one complex depends on the local concentration of the other. 
At the linear level, the antisymmetric mode is the first to lose stability when $\mathcal{V}'(\bar{c})>0$ (Section~\ref{PCP_dispersion}). Fully formed puncta inherit this same criterion: a co-located pair segregates if and only if $\mathcal{V}'>0$ (Section~\ref{PCP_polarity}). 
This sign-definite, competitive cross-interaction is exactly the destabilising role attributed to Prickle, which has been proposed to promote the removal of complexes of opposite orientation~\citep{warrington2017dual, cho2015clustering}. 
In the model, then, the mass-redistribution machinery selects \emph{how many} puncta form while the sign of the cross-coupling selects \emph{whether they sort}; clustering and sorting are controlled by separate ingredients. 
We stress once more that what the model delivers is mutual exclusion of the two orientations along a single contact. 
Cell-scale planar polarity requires in addition that the contacts of a cell be coupled and that some bias select which orientation prevails at which edge; both lie outside the present model, and we return to this in the outlook below.
 
\paragraph{Stable versus metastable coexistence.} 
Which of the multi-punctum states above are genuinely stable, and which only long-lived? 
The distinction turns on whether two puncta draw on the same conserved reservoir. 
Two puncta of the \emph{same} orientation share the monomer pool of a single triplet, and the competition eigenvalue $\lambda_{\mathrm{comp}}=\alpha\bar{a}^{3}D/(3\ell)$ of \eqref{PCP_rate_explicit} is positive for all parameters: the antisymmetric (grow-one, shrink-the-other) mode is always unstable and the configuration coarsens to a single spike. 
Asymmetric diffusion only rescales this rate to the harmonic mean \eqref{PCP_harmonic}, and the spectral analysis of Section~\ref{PCP_asym} confirms that the relevant amplitude mode stays real, so no parameter choice stabilises a same-orientation array. 
Because $\lambda_{\mathrm{comp}}\propto1/\ell$, coarsening is slow on large domains and many spikes coexist for long times -- the metastable states of Figure~\ref{AB_onset_numerics} -- but the only true attractor with same-orientation puncta is the single spike. 

The wave-pinning mesas reach the same endpoint, though by a weaker interaction. 
Two like-oriented fronts are not held apart by any conservation-protected mode, so on a conserved domain a pair of mesas drawing on a single reservoir again has only the single mesa as a true attractor. 
Neighbouring mesa interfaces communicate through the exponentially small overlap of their tails rather than through the algebraic, screening-length coupling of spikes, so their coarsening is exponentially slow in the inter-front separation rather than algebraic. 
Like-oriented mesas therefore coarsen towards a single front, but at a rate set by this exponentially-weak tail interaction, and are correspondingly more strongly metastable than spikes. 
The qualitative conclusion -- that a single same-orientation punctum is the only stable state -- is unchanged, even though the approach to it is not at the spike rate.
 
\paragraph{Coexistence across orientations.} 
Puncta of \emph{opposite} orientation are different. 
The two triplets are independently conserved and draw on disjoint monomer pools, so a $c$- and a $c^{\dagger}$-punctum share no competition mode and cannot annihilate one another; what remains is only their relative arrangement, governed by $\mathcal{V}'$. 
At $\mathcal{V}'=0$ the pair coexists, but their separation is a neutral mode fixed by the initial data. 
At $\mathcal{V}'>0$ the co-located pair is unstable and the \emph{segregated} pair is the stable endpoint: the cross-repulsion \eqref{PCP_drift} restores the separation against the merge mode (Section~\ref{PCP_polarity}). 
This sorted pair is the one configuration in the model with two spatially distinct, asymptotically stable puncta. 
Each conserved triplet thus supports exactly one stable punctum, and with two triplets the maximum genuinely stable count is two -- one of each orientation -- pinned in space only when $\mathcal{V}'>0$.
 
\paragraph{Stable arrays require non-conservative ingredients.} The coexistence analysis is a positive result with a sharp boundary. 
What mass conservation \emph{permits} is sorting: each conserved reservoir relaxes to a single punctum, so the two reservoirs together support a stable, spatially pinned pair once $\mathcal{V}'>0$ (Sections~\ref{PCP_competition}, \ref{PCP_polarity}). 
What it \emph{forbids} is stable multiplicity. 
By the same competition argument -- the eigenvalue \eqref{PCP_rate_explicit} is positive for every parameter choice -- a regular array of many like-oriented puncta, of the kind seen at core-PCP junctions, is not a stable state of these kinetics but a metastable transient that must eventually coarsen. 
Recent molecular counting of core-protein clusters has been interpreted as favouring size-independent growth and decay over coarsening~\citep{nissen2026cluster}; the interpretation of those measurements, and its bearing on the coarsening predicted here, is taken up elsewhere. 
We regard this as a structural property of mass conservation rather than a shortcoming of the present treatment. 
Stable $N$-spike solutions are known to require production--decay (Gierer--Meinhardt-type) turnover, which breaks conservation \citep{gierer1972theory, iron2001stability}, with an externally imposed length scale or sequential deposition by domain growth as alternative routes; the array-stabilising ingredient therefore lies, by construction, outside the conserved kinetics studied here. 
Confining the model to the conserved regime is a deliberate choice that isolates sorting in its simplest setting. 
The most natural next step is to identify the \emph{minimal} departure from conservation that restores a stable multi-punctum array while preserving the sorting mechanism of Section~\ref{PCP_polarity}.
 
\paragraph{Persistence and the absence of blinking.} 
The conserved kinetics also constrain the dynamics of an established punctum, which is a stable, damped focus rather than an oscillator. 
Sustained ``blinking'' would require a Hopf bifurcation, and the conserved structure forbids one (Section~\ref{PCP_asym}): it would need non-conservative turnover or an explicit feedback delay. 
This is consistent with the experimental picture of core-protein puncta as persistent structures of reduced turnover~\citep{strutt2011dynamics, strutt2016robust} rather than transiently flickering ones.
 
\paragraph{Relation to pattern-forming mechanisms.} 
Mathematically, the model sits at the confluence of two mechanisms. 
The combination of bistable well-mixed kinetics, mass conservation, and faster diffusion of the monomers than of the complex is precisely the wave-pinning setting of \citet{mori2008wave}, and the instability is, in the terminology of \citet{brauns2020phase}, a mass-redistribution instability; the spike regime, by contrast, inherits the singular structure of activator--substrate spikes~\citep{gierer1972theory, iron2001stability}, adapted to the near-shadow, mass-conserving limit~\citep{mckay2012stability}. 
Read against the substantial literature on mass-conserving systems~\citep{brauns2021wavelength, frey2026pattern}, most of the qualitative phenomenology reported here is what that theory would lead one to expect: the long-wavelength onset, the single-peak attractor, the uninterrupted coarsening, and the kinetic rather than wavelength-based selection of number. 
The distinctive feature is the pair of \emph{independently} conserved reservoirs demanded by a \emph{trans}-complex, and it enters in two ways that have no single-species analogue. 
Quantitatively, it makes competition rate-limited by the harmonic mean of the two mobilities (Section~\ref{PCP_asym}), a law with no single-reservoir counterpart. 
Structurally, it supplies two reservoirs to be sorted \emph{between}, and hence the antisymmetric mode of Section~\ref{PCP_polarity}: a one-species clustering model has nothing to sort. 
The homodimer of Section~\ref{PCP_homodimer} illustrates this: it clusters by the same mechanism but, being homophilic, admits no sorting mode at all.
 
\paragraph{Limitations and outlook.} 
Several simplifications temper these conclusions. 
The junction is treated as one-dimensional; the biochemistry is reduced to two complex species with feedback entering only through the binding and unbinding rates $\mathcal{K}$ and $\mathcal{V}$; and $\mathcal{V}$ is a scalar caricature of the phosphorylation- and endocytosis-mediated antagonism among the core proteins~\citep{strutt2019reciprocal}.
Processes known to shape planar polarity \emph{in vivo} -- directed vesicular transport, recycling and turnover of complex components at the membrane, and the phosphorylation-dependent modulation of binding and stability~\citep{strutt2019reciprocal} -- enter here only implicitly, through the effective rates $\mathcal{K}$ and $\mathcal{V}$. We also do not consider the interaction between the two pathways, core and Fat--Dachsous, that together pattern the tissue.
The idealised two-species description adopted here is a deliberate choice that trades molecular detail for analytical insight, and isolates the generic ingredients of clustering and sorting. 
Two extensions are natural. 
The first is to develop a more biophysically mechanistic version of the model with explicit molecular players, in which the feedback rates $\mathcal{K}$ and $\mathcal{V}$ are derived from the underlying phosphorylation and endocytosis kinetics rather than imposed. 
The second is to ask how this molecular-scale self-organisation feeds up to cell-scale polarity: the local organisation of proteins into puncta at a cell-cell junction does not, by itself, explain the self-organisation of coordinated cell polarisation, which additionally requires the symmetry breaking that underlies whole-cell polarity and the cell--cell coupling that aligns it between neighbours~\citep{goodrich2011principles}.

\section*{Acknowledgments}
This work was supported by UKRI-EPSRC, through grant EP/W024144/1 to A.G.F.\ and grant EP/R014604/1, and by a PhD scholarship to E.A.\ from Umm Al-Qura University funded by the Government of Saudi Arabia. 
The authors would like to thank the Isaac Newton Institute for Mathematical Sciences, Cambridge, for support and hospitality during the programme `Mathematics of movement: an interdisciplinary approach to mutual challenges in animal ecology and cell biology', where work on this paper was undertaken. 
Generative AI tools were used to assist in the development of the figure-generation code and in editing the manuscript text. 
The authors verified all code, derivations, and text; the code is publicly available at \url{https://github.com/AlexFletcher/puncta} to enable reproducibility. 
The authors assume responsibility for all content. 
For the purpose of open access, the author has applied a Creative Commons Attribution (CC BY) licence to any Author Accepted Manuscript version arising.
 
\clearpage
\bibliographystyle{unsrtnat}
\bibliography{references}

\end{document}